\documentclass[12pt,a4paper]{article}

\usepackage{lineno}  % for line numbering during review

\usepackage{graphicx}  % to include figures (can also use other packages)

\usepackage{xspace}
\usepackage{color}
\usepackage{colortbl}

\usepackage{amsmath}
\usepackage{ifthen} % for conditional statements

\newboolean{articletitles}
\setboolean{articletitles}{true}

\usepackage{overpic}

\newboolean{pdflatex}
\setboolean{pdflatex}{true} % use this if using non-eps figures
\usepackage{rotating} 
\newboolean{uprightparticles}
\setboolean{uprightparticles}{false} %Set to true to get roman particle symbols
\usepackage{amssymb}
\usepackage{amsfonts}
\usepackage{upgreek}
\usepackage{lhcb-symbols-def}

\usepackage{hyperref}
\usepackage[all]{hypcap}

\usepackage{cite}
\usepackage{mciteplus}

\begin{document}

%%%%%%%%%%%%%%%%%%%%%%%%%
%%%%% Title     %%%%%%%%%
%%%%%%%%%%%%%%%%%%%%%%%%%

% %%%%%%% CHOOSE --------
%  Choose the right title template or customize existing one if your type is still missing
%\input{title-LHCb-CONF}
\renewcommand{\thefootnote}{\fnsymbol{footnote}}
\setcounter{footnote}{1}
% $Id: title-LHCb-PAPER.tex 16117 2012-02-21 14:00:45Z uegede $
% ===============================================================================
% Purpose: LHCb-PAPER journal paper title page template
% Author: 
% Created on: 2010-09-25
% ===============================================================================

%%%%%%%%%%%%%%%%%%%%%%%%%
%%%%%  TITLE PAGE  %%%%%%
%%%%%%%%%%%%%%%%%%%%%%%%%
\begin{titlepage}
\pagenumbering{roman}

% Header ---------------------------------------------------
\vspace*{-1.5cm}
\centerline{\large EUROPEAN ORGANIZATION FOR NUCLEAR RESEARCH (CERN)}
\vspace*{1.5cm}
\hspace*{-0.5cm}
\begin{tabular*}{\linewidth}{lc@{\extracolsep{\fill}}r}
\ifthenelse{\boolean{pdflatex}}% Logo format choice
{\vspace*{-2.7cm}\mbox{\!\!\!\includegraphics[width=.14\textwidth]{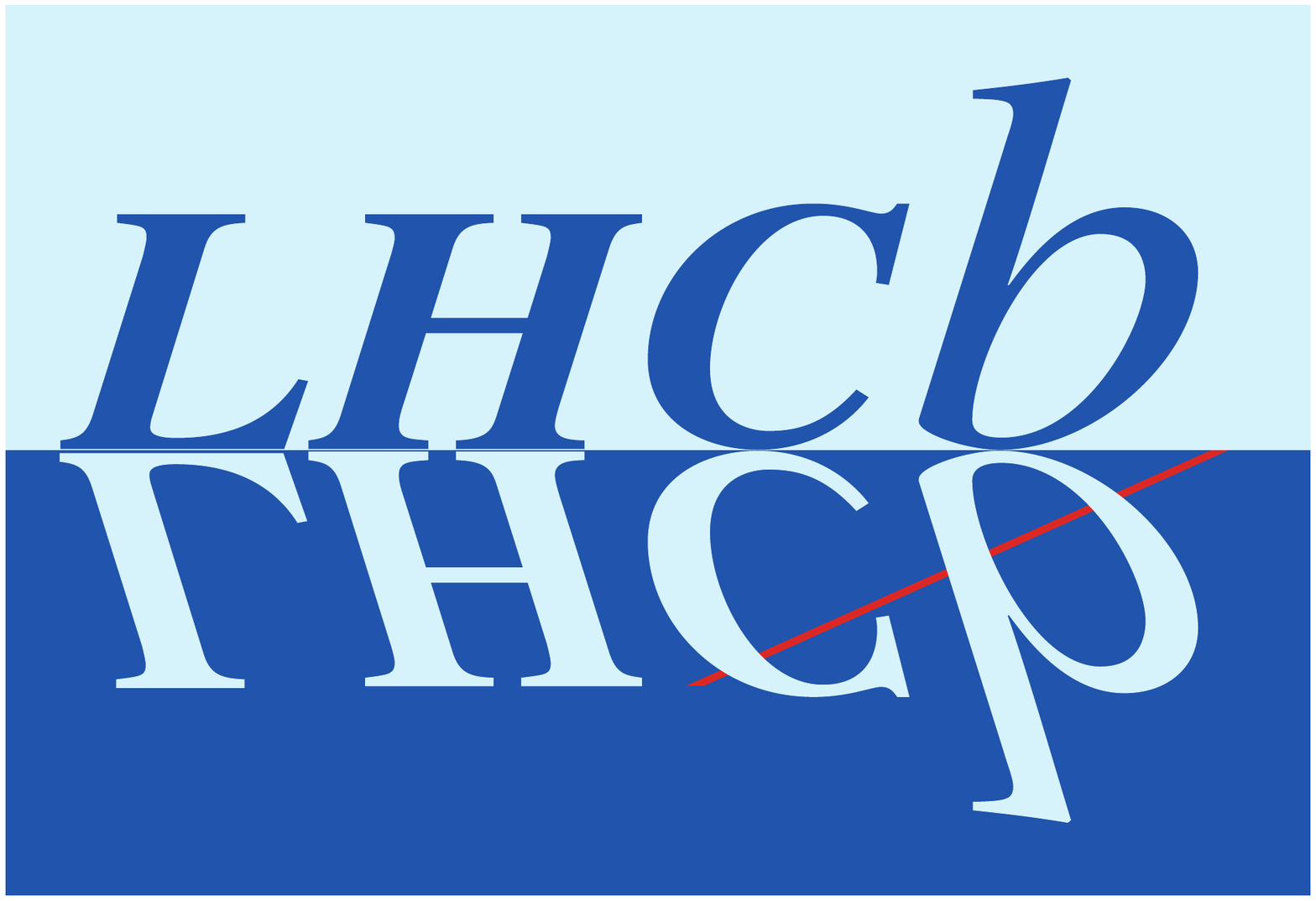}} & &}%
{\vspace*{-1.2cm}\mbox{\!\!\!\includegraphics[width=.12\textwidth]{lhcb-logo.eps}} & &}%
\\
 & & LHCb-PAPER-2011-027   \\  % ID 
 & & CERN-PH-EP-2012-039 \\  % ID 
 & & \today \\ % Date - Can also hardwire e.g.: 23 March 2010
 & & \\
\end{tabular*}

\vspace*{4.0cm}

% Title --------------------------------------------------
{\bf\boldmath\huge
\begin{center}
Opposite-side flavour tagging of $B$ mesons at the LHCb experiment
\end{center}
}

\vspace*{2.0cm}

% Authors -------------------------------------------------
\begin{center}
The LHCb collaboration
\footnote{Authors are listed on the following pages.}
\end{center}

\vspace{\fill}

% Abstract -----------------------------------------------
\begin{abstract}
  \noindent
The calibration and performance of the opposite-side 
   flavour tagging algorithms used for the measurements of 
time-dependent asymmetries at the LHCb experiment are described.
 The algorithms have been developed using simulated events and
optimized and calibrated with  \bplus, \BdJKst\ and 
\dstarmunu decay modes with 0.37 \invfb of data collected in $pp$ collisions 
at $\sqrt{s} =$ 7\tev during the 2011 physics run.
The opposite-side tagging power is determined in the 
$B^+\to$~$J/\psi K^+$
channel to be (2.10$\pm$0.08$\pm$0.24)\%, where the first uncertainty 
is statistical and the second is systematic.

\end{abstract}

\vspace*{1.0cm}
\begin{center}
Submitted to Eur. Phys. J. C
\end{center}

\vspace*{2.0cm}
\vspace{\fill}

\end{titlepage}

%%%%%%%%%%%%%%%%%%%%%%%%%%%%%%%%
%%%%%  EOD OF TITLE PAGE  %%%%%%
%%%%%%%%%%%%%%%%%%%%%%%%%%%%%%%%

%  empty page follows the title page ----
\newpage
\setcounter{page}{2}
\mbox{~}
\newpage

% Author List ----------------------------
% \documentclass[a4paper]{article}
% \setlength{\oddsidemargin}{0cm}
% \setlength{\evensidemargin}{0cm}
% \setlength{\textwidth}{16.5cm}
% \setlength{\parindent}{0cm}
% \begin{document}
\centerline{\large\bf The LHCb collaboration}
\begin{flushleft}
\small
%{\Large  LHCb Collaboration }----- official authorship list}\\[4ex]
% valid for date: 4. Dec. 2011\\
% used for paper: Flavour tagging (LHCb-PAPER-2011-027)\\[4ex]
% collaborators included, who did not leave before 4. Dec. 2010\\
%                            and who joined before 4. Jun. 2011\\[2ex]
% {\small today is 5. Feb. 2012}\\[4ex]
% %-- 
% %-- LHCb Authorlist, Status of 4. Dec. 2011
% %-- 
R.~Aaij$^{38}$, 
C.~Abellan~Beteta$^{33,n}$, 
B.~Adeva$^{34}$, 
M.~Adinolfi$^{43}$, 
C.~Adrover$^{6}$, 
A.~Affolder$^{49}$, 
Z.~Ajaltouni$^{5}$, 
J.~Albrecht$^{35}$, 
F.~Alessio$^{35}$, 
M.~Alexander$^{48}$
% ...

{\footnotesize \it
% ...
$ ^{5}$Clermont Universit\'{e}, Universit\'{e} Blaise Pascal, CNRS/IN2P3, LPC, Clermont-Ferrand, France\\
$ ^{6}$CPPM, Aix-Marseille Universit\'{e}, CNRS/IN2P3, Marseille, France\\
% ...
$ ^{33}$Universitat de Barcelona, Barcelona, Spain\\
$ ^{34}$Universidad de Santiago de Compostela, Santiago de Compostela, Spain\\
$ ^{35}$European Organization for Nuclear Research (CERN), Geneva, Switzerland\\
% ...
$ ^{38}$Nikhef National Institute for Subatomic Physics, Amsterdam, The Netherlands\\
% ...
$ ^{43}$H.H. Wills Physics Laboratory, University of Bristol, Bristol, United Kingdom\\
% ...
$ ^{48}$School of Physics and Astronomy, University of Glasgow, Glasgow, United Kingdom\\
$ ^{49}$Oliver Lodge Laboratory, University of Liverpool, Liverpool, United Kingdom\\
% ...
$ ^{n}$LIFAELS, La Salle, Universitat Ramon Llull, Barcelona, Spain\\
% ...
}
% \bigskip
% ---- LHCb Authorlist, Status 4. Dec. 2011
% ---- Number of Authors = 596
% ---- 
\end{flushleft}
%\end{document}

\newpage

\renewcommand{\thefootnote}{\arabic{footnote}}
\setcounter{footnote}{0}

% %%%%%%%%%%%%% ---------

%%%%%%%%%%%%%%%%%%%%%%%%%%%%%%%%
%%%%%  Table of Content   %%%%%%
%%%%%%%%%%%%%%%%%%%%%%%%%%%%%%%%
%%%% Uncomment next 2 lines if desired
%\tableofcontents

%%%%%%%%%%%%%%%%%%%%%%%%%
%%%%% Main text %%%%%%%%%
%%%%%%%%%%%%%%%%%%%%%%%%%

\pagestyle{plain} % restore page numbers for the main text
\setcounter{page}{1}
\pagenumbering{arabic}

% %%%%%%% CHOOSE --------
%% ----------------------------------
%% Line numbering on the left margin 
%% ----------------------------------
%% Uncomment during review phase. 
%% Comment it out before a final submission.
%\linenumbers
%% --------------------------------
% %%%%%%%%%%%%% ---------

%  You can include short sections directly in the main tex file. 
%  However, for larger papers it is desirable to split 
%  the text into several semiautonomous files, which can be revised independently. 
%  This is especially useful when developing a document in collaboration with several people, 
%  since then different parts can be edited independently. 
%  This type of file organization is shown here. 
% 
%%%%%%%%%%%%%%%%%%%%%%%%%%%%%%%%%%%%%%%%%%%
\section{Introduction} \label{sec:introduction}
The identification of the flavour of reconstructed \Bd and \Bs mesons 
at production is necessary for the measurements of oscillations 
and time-dependent \CP\ asymmetries.
This procedure is known as {\it flavour tagging} and is performed 
at \lhcb by means of several algorithms.

Opposite-side (OS) tagging algorithms rely on the pair production
of $b$ and $\bar b$ quarks and infer the flavour of a given $B$ meson 
(signal $B$) from the identification of the flavour of the other 
$b$ hadron\footnote{Unless explicitly stated, charge conjugate modes are 
always included throughout this paper.} (tagging $B$).
The algorithms use the charge of the lepton ($\mu$, $e$) 
from semileptonic $b$ decays, the charge of the kaon from the $b\to c\to s$ 
decay chain or the charge of the inclusive secondary vertex 
%($Q_{\mathrm{vtx}}$) 
reconstructed from $b$-hadron decay products.
All these methods have an intrinsic dilution on the tagging decision, 
for example due to the 
possibility of flavour oscillations of the tagging $B$.
This paper describes the optimization and calibration of the OS 
tagging algorithms which are performed with the data used for the 
first measurements performed 
by LHCb on \Bs\ mixing and time-dependent 
\CP\ violation~\cite{DMs_2010,phis_f0,phis_phi}.

Additional tagging power can be derived from same-side tagging algorithms
which determine the flavour of the signal $B$ by exploiting its correlation  
with particles produced in the hadronization process. The use of these 
algorithms at LHCb will be described in a forthcoming publication.
The use of flavour tagging in previous experiments at hadron colliders 
is described in Refs.~\cite{tagging_D0,tagging_CDF}.

The sensitivity of a measured \CP\ asymmetry is directly related to the 
effective tagging efficiency \effeff, or tagging power.
The tagging power represents the effective statistical reduction of the 
sample size, and is defined as
\begin{equation}\label{eq:effeff}
\effeff  = \effD = \efftag ,
\end{equation}
\noindent 
where 
\etag\ is the tagging efficiency, \mistag\ is
the mistag fraction and $\Dilu$  is the dilution. The tagging efficiency
and the mistag fraction are defined as 
\begin{equation}\label{eq:tageffdef}
\etag = \frac{R + W}{R + W + U} \qquad {\rm and} ~~~~~ \omega = \frac{W}{R + W},
\end{equation}
\noindent where $R$, $W$, $U$ are the number of correctly tagged,
incorrectly tagged and untagged events, respectively.

The  mistag fraction can be measured in data using flavour-specific  
decay channels, i.e. those decays where the final state particles 
uniquely define the quark/antiquark content of the signal $B$.
In this paper,  the decay channels \bplus, \bkstar and \dstarmunu are used.
For charged mesons, the mistag fraction is obtained by directly comparing the 
tagging decision with the flavour of the signal $B$, while for neutral 
mesons it is obtained by fitting the \Bd flavour oscillation as a 
function of the decay time.

The probability of a given tag decision to be correct is estimated 
from the kinematic properties of the tagging particle and the event itself
by means of a  neural network trained on Monte Carlo (MC) simulated 
events to identify the correct flavour of the signal $B$.
When more than one tagging algorithm gives a response for an event, 
the probabilities provided by each algorithm are combined into a single 
probability and the decisions are combined into a single decision.
The combined probability can be exploited on an
event-by-event basis to assign larger weights to events with low mistag
probability and thus to increase the overall significance of an asymmetry 
measurement. 
In order to get the best combination and a reliable estimate 
of the event weight, the calculated probabilities are calibrated on data.
The default calibration parameters are extracted 
from the \bplus channel.
The other two flavour-specific channels  are used to perform independent 
checks of the calibration procedure.

\section{The LHCb detector and the data sample}\label{sec:Detector}

The \lhcb detector~\cite{LHCb} is a single-arm forward spectrometer which 
measures \CP\ violation and rare decays of hadrons containing $b$ and $c$ 
quarks.  
A vertex detector (VELO) determines with high precision the positions of the 
primary and secondary vertices as well as the impact parameter (${\rm IP}$) 
of the reconstructed tracks with respect to the primary vertex. 
The tracking system also includes a silicon strip detector located in front
of a dipole magnet with integrated field about  4\,Tm, and a combination of 
silicon strip detectors and straw drift chambers placed behind the magnet.  
Charged hadron identification is achieved through two ring-imaging Cherenkov 
(RICH) detectors. The calorimeter system consists of a preshower detector, 
a scintillator pad
detector, an electromagnetic calorimeter and a hadronic calorimeter. 
It identifies high transverse energy hadron, electron and photon candidates 
and provides information
for the trigger. Five muon stations composed of multi-wire proportional 
chambers and triple-GEMs (gas electron multipliers) provide fast information 
for the trigger and muon identification capability.

The \lhcb trigger consists of two levels. The first, hardware-based, level 
selects leptons and hadrons with high transverse momentum, using the 
calorimeters and the muon detectors. 
The hardware trigger is followed by a software High Level Trigger (HLT), 
subdivided into two stages that use the information from all parts of the 
detector. 
The first stage performs a partial reconstruction of the event, 
reducing the rate further and allowing the next stage to fully reconstruct and
to select the events for storage up to a rate of 3\,kHz~\cite{lhcb-2011-016}.

The majority of the events considered in this paper were triggered by a single 
hadron or muon track with large momentum, transverse momentum and $\rm IP$. 
In the HLT, the channels with a \jpsi meson in the final state were selected  
by a dedicated di-muon decision that does not apply any requirement on 
the $\rm IP$ of the muons.  

The data used in this paper were taken between March and June 2011  
and correspond to an integrated luminosity of 0.37 \invfb. 
The polarity of the LHCb magnet was reversed several times during the 
data taking period in order to minimize systematic biases due to
possible detector asymmetries.

\section{Flavour tagging algorithms}\label{sec:FlavourTaggingAlgo}

Opposite-side tagging uses the identification of electrons, muons or kaons 
that are attributed to the other $b$ hadron in the event.
It also uses the charge of tracks consistent with 
coming from a secondary vertex not associated with either the primary or the 
signal $B$ vertex. 
These taggers are called electron, muon, kaon and vertex charge taggers, 
respectively.
The tagging algorithms were developed and studied 
using simulated events~\cite{lhcb-2007-058}. 
Subsequently, the criteria to select the tagging particles and to reconstruct 
the vertex charge are re-tuned, using the \bplus and the \dstarmunu 
control channels.
An iterative procedure is used to find the selection criteria which 
maximize the tagging power \effeff.

Only charged particles reconstructed with a good quality of the track fit 
are used.
In order to reject poorly reconstructed tracks, the track is required to have
a polar angle with respect to the beamline larger than 12~\mrad and a 
momentum larger than 2~\gevc.
Moreover, in order to avoid possible duplications of the signal tracks, 
the selected particles are required to be outside a cone of 5~\mrad formed 
around any daughter of the signal $B$.
To reject tracks coming from other primary interactions in the same 
bunch crossing, the impact parameter significance with respect to
these pile-up ($\rm PU$) vertices, $\rm{IP_{PU}}/\sigma_{\rm IP_{PU}} > 3$,
is required.

\subsection{Single-particle taggers}

The tagging particles are selected exploiting 
the properties of the  $b$-hadron decay. 
A large impact parameter significance with respect to the primary vertex 
($\rm{IP}/\sigma_{\rm IP}$)
and a large transverse momentum~\pt are required. 
Furthermore, particle identification cuts are used to define each tagger based 
on the information from the RICH, calorimeter and muon systems.
For this purpose, the differences between the logarithm of the likelihood for 
the muon, electron, kaon or proton and the pion hypotheses (referred as 
${\rm DLL}_{\mu-\pion}$, ${\rm DLL}_{e-\pion}$, ${\rm DLL}_{K-\pion}$ 
and ${\rm DLL}_{p-\pion}$) are used.
The detailed list of selection criteria is reported in Table~\ref{tab:Tagcuts}. 
Additional criteria are used to identify the leptons.
Muons are required not to share hits in the muon chambers with other tracks,
in order to avoid mis-identification of tracks which are close 
to the real muon.
Electrons are required to be below a certain threshold in the ionization 
charge deposited in the silicon layers of the VELO, 
in order to reduce the number of 
candidates coming from photon conversions close to the interaction point.
An additional cut on the ratio of the particle energy $E$ as measured in 
the electromagnetic calorimeter and the momentum $p$ of the candidate electron
 measured with the tracking system, $E/p > 0.6$, is applied.
 
In the case of multiple candidates from the same tagging algorithm, the 
single-particle tagger with the highest \pt is chosen and its charge is used 
to define the flavour of the signal $B$.

%TTTTTTTTTTTTTTTTTTTTTTTTTTTTTTTTTTTTTTTTTTTTTTTTTTTTT
\begin{table}
\begin{center}
\caption{\small{Selection criteria for the OS muon, electron and kaon taggers.}}
\begin{tabular} {c | c | c | c | l| c }
\hline 
Tagger& min \pt & min \ptot & min (${\rm IP}/\sigma_{\rm IP}$) 
&Particle identification& min (${\rm IP_{PU}}/\sigma_{\rm IP_{PU}}$)\\
     &[\gevc] & [\gevc]& & \;\;\;\;\;\;\;\;\;\; cuts &  \\ \hline
$\mu$& 1.2    & 2.0     & -   & ${\rm DLL}_{\mu-\pion} >2.5 $ & 3.0  \\ \hline
$e$    & 1.0      & 2.0     & 2.0   & ${\rm DLL}_{ e-\pion} > 4.0$     & 3.0  \\ \hline
$K$    & 0.8    & 5.9 & 4.0& ${\rm DLL}_{K-\pion}> 6.5$   & 4.7 \\
     &        &       &     & ${\rm DLL}_{K-p} >-3.5$ &  \\ \hline
\end{tabular}
\label{tab:Tagcuts}
\end{center}
\end{table}
%TTTTTTTTTTTTTTTTTTTTTTTTTTTTTTTTTTTTTTTTTTTTTTTTTTTTT

\subsection{Vertex charge tagger}
The vertex charge tagger is based on the inclusive reconstruction of a 
secondary vertex corresponding to the decay of the tagging $B$.
The vertex reconstruction consists of building a composite candidate 
from two tracks with a transverse momentum \mbox{$\pt >0.15~\gevc$} and
$\rm{IP}/\sigma_{\rm IP}>2.5$. The pion mass is attributed to the tracks.
Moreover, good quality of the vertex reconstruction is required 
and track pairs with an invariant mass compatible with a \KS meson are 
excluded. 
For each reconstructed candidate 
the probability that it originates from a $b$-hadron decay is estimated
from the quality of the vertex fit as well as from the geometric and kinematic 
properties.
Among the possible candidates the one with the highest probability is used.
Tracks that are compatible with coming from the two track vertex but
do not originate from the primary vertex are added to form the final candidate.
Additional requirements are applied to the tracks asspociated to the 
reconstructed secondary vertex:
total momentum~$> 10$~\gevc, total \pt $> 1.5$~\gevc , total invariant mass 
$> 0.5$~\gevcc and the sum of $\rm{IP}/\sigma_{\rm IP}$ 
of all tracks $> 10$.

Finally, the charge of the tagging $B$ is calculated as
the sum of the charges $Q_i$ of all the tracks associated to the vertex, 
weighted with their transverse momentum to the power $\kappa$ 
\begin{equation}
Q_{\rm vtx} = \frac{\Sigma_i Q_i p^{\kappa}_{\rm Ti} }{\Sigma_ip^{\kappa}_{\rm Ti} },
\end{equation}
\noindent where the value $\kappa=0.4$ optimizes the tagging power.
Events with $|Q_{\rm vtx}| < 0.275$ 
are rejected as untagged.

%%%%%%%%%%%
\subsection{Mistag probabilities and combination of taggers}
\label{sec:CombTaggers}
%%%%%%%%%%%

For each tagger $i$, the probability $\eta_i$ of the tag decision to be wrong
is estimated by using properties of the  tagger and of the event 
itself. 
This mistag probability is evaluated by means of a neural network trained 
on simulated \BuJK\ events to identify the correct flavour of the signal 
$B$ and subsequently calibrated on data 
as explained in Sect.~\ref{sec:tagging_calibration}. 

The inputs to each of the neural networks are the signal $B$ transverse 
momentum, the number of pile-up vertices, the number of tracks preselected
as tagging candidates and various geometrical and kinematic 
properties of the tagging particle (\ptot, \pt and ${\rm IP}/\sigma_{\rm IP}$ 
of the particle), or of the tracks associated to the secondary vertex 
(the average values of \pt, of $\rm IP$, the reconstructed 
invariant mass and the absolute value of the vertex charge).

If there is more than one tagger available per event, the decisions provided 
by all available taggers are combined into a final decision on the initial 
flavour of the signal $B$.
The combined probability $P(b)$ that the meson contains a $b$-quark is 
calculated as
\begin{equation}\label{eq:mistagProb}
P(b) = \frac{p(b)}{p(b)+p(\bar{b})}, \qquad \quad P(\bar{b})=1-P(b),
\end{equation}
where
\begin{equation} \label{eq:probb}
p(b) = \prod_i \left( \frac{1+d_i}{2} - d_i (1-\eta_i) \right), 
\qquad \quad p(\bar{b}) = \prod_i \left (\frac{1-d_i}{2} + d_i (1- \eta_i) \right). 
\end{equation}
\noindent Here, $d_i$ is the decision taken by the $i$-th tagger
based on the charge of the particle with the convention
$d_i=1(-1)$ for the signal $B$ containing a $\bar b$($b$) quark and $\eta_i$ the corresponding predicted mistag probability.
The combined tagging decision and the corresponding mistag probability 
are $d=-1$ and $\eta = 1 - P(b)$ if $P(b)>P(\bar b)$, otherwise $d=+1$ 
and $\eta = 1 - P(\bar b)$.

The contribution of taggers with a poor tagging power is limited by requiring
the mistag probabilities of the kaon and the vertex charge
to be less than 0.46.

Due to the correlation among taggers, which is neglected in 
Eq.~\ref{eq:probb}, the combined probability is slightly 
overestimated.
The largest correlation occurs between the vertex charge 
tagger and the other OS taggers, since the secondary vertex may 
include one of these particles.
To correct for this overestimation, the combined OS probability is
calibrated on data, as described in Sect.~\ref{sec:tagging_calibration}.

%%%%%%%%%%%
\section{Control channels}
%%%%%%%%%%%
The flavour-specific \B decay modes \bplus, \bkstar and  \dstarmunu
are used for the tagging analysis.
All three channels are useful to optimize the performance of the OS tagging 
algorithm and to calibrate the mistag probability.
The first two channels are chosen 
as representative control channels for the decays \BsToJPsiPhi
and {\decay{\Bs}{\jpsi f_0}}, which are used for the measurement 
of the \Bs mixing phase $\phi_s$~\cite{phis_f0,phis_phi}, 
and the last channel allows detailed studies given
the high event yield of the semileptonic decay mode.
All \B decay modes with a \jpsi\ meson in the final state share 
the same trigger selection and common offline selection criteria,
which ensures a similar performance of the tagging algorithms.
Two trigger selections are considered, with or without requirements 
on the $\rm IP$ of the tracks. They are labelled ``lifetime biased'' 
and ``lifetime unbiased'' respectively.

%%%%%%%%%%
\subsection{ Analysis of the  $\boldsymbol{B^+ \to J/ \psi K^+}$ channel }
\label{sec:bplus}
%%%%%%%%%%%

The \bplus candidates are selected by combining $J/\psi \to \mu^+ \mu^-$ 
 and $K^+$ candidates. 
The \jpsi mesons are selected by combining two muons with transverse momenta 
$\pt>$~0.5~\gevc that form a common vertex of good quality 
and have an invariant mass in the range $3030-3150$~\mevcc. 
The $K^+$ candidates are required to have transverse momenta $\pt > 1$~\gevc 
and momenta $\ptot >10$~\gevc and to form a common vertex of good quality with 
the $J/\psi$ candidate with a resulting invariant
mass in a window $\pm90$~\mevcc around the \Bp mass. 
Additional requirements on the particle identification of muons and kaons 
are applied to suppress the background contamination.
To enhance the sample of signal events and reduce the dominant background 
contamination from prompt $J/\psi$ mesons combined with random kaons, 
only the events
 with a reconstructed decay time of the \Bu candidate $t>0.3$\ps are selected. 
The decay time $t$ and the invariant mass $m$ of the \Bu meson
are extracted from a vertex fit that includes a constraint on the associated 
primary vertex, and a constraint on the \jpsi mass for the evaluation of the $J/\psi K$ invariant mass.
In case of multiple $B$ candidates per event, only the one with the smallest 
vertex fit $\chi^2$ is considered.  

\begin{figure}[ht]
  \begin{center}
    \includegraphics[angle=0, width=0.50\textwidth]{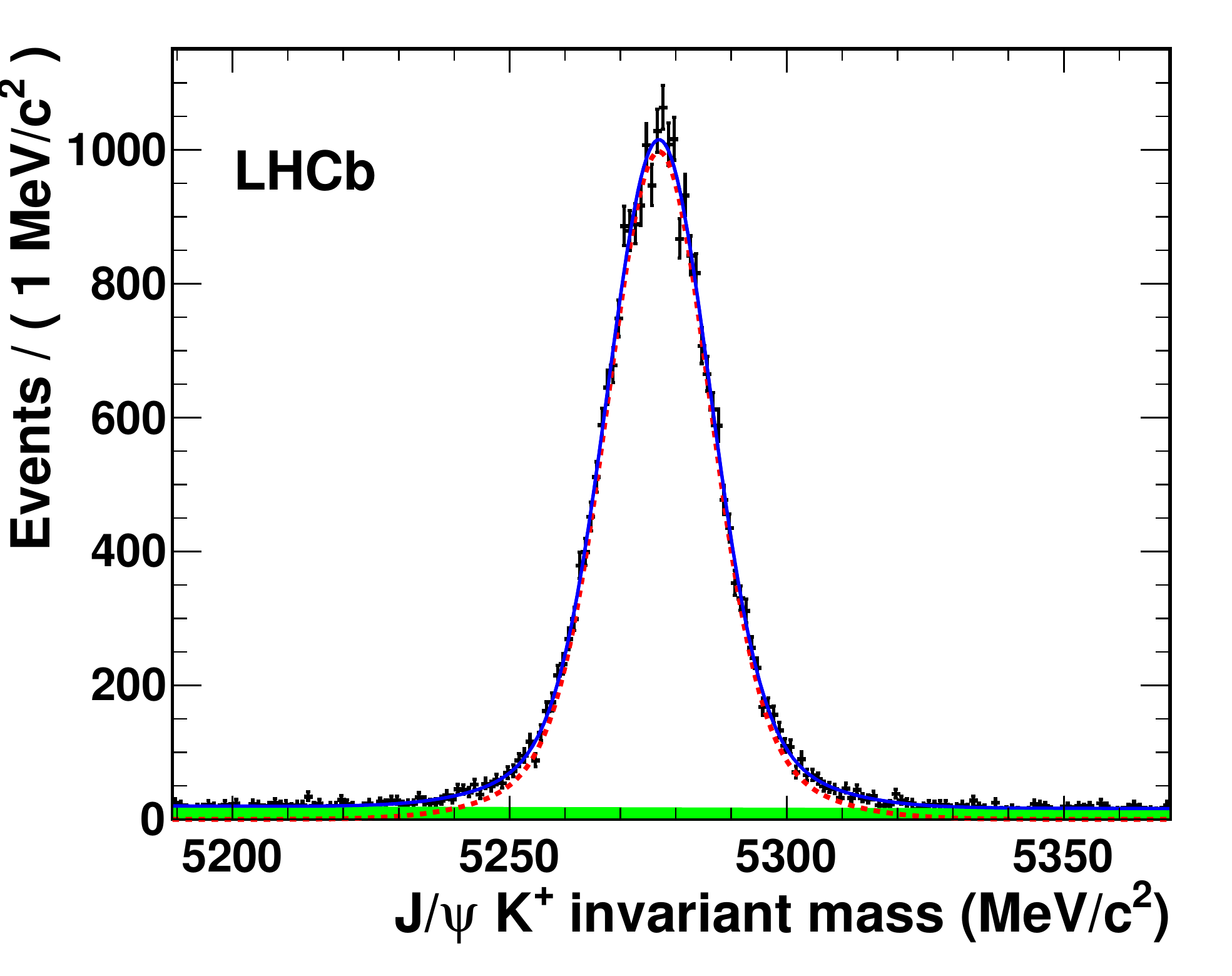}
     \end{center}
     \caption{\small{Mass distribution of OS tagged \bplus events. Black points
 are data, the solid  blue line, red dotted line and green area are the overall
fit, the signal and the background components, respectively.
}}
\label{fig:mass_bplus}
\end{figure}

The signal events are statistically  disentangled from the background, which is dominated by partially reconstructed 
$b$-hadron decays to $J/\psi K^+ X$ (where $X$ represents any other particle 
in the decay), by 
means of an unbinned maximum likelihood fit to the reconstructed \Bu mass and decay time.
In total $\sim 85\,000$ 
signal events  are selected with a background to signal ratio $B/S\sim0.035$, 
calculated in a window of $\pm40$~\mevcc centred around the \Bu mass.
The mass fit model is based on  a double Gaussian distribution peaking at the \Bu mass for the signal and an exponential distribution for the background. 
The time distributions of both the signal and the background are assumed to be exponential, with separate decay constants. 
The fraction of right, wrong or untagged events in the sample is determined
according to a probability density function (PDF), ${\cal P}(r)$, 
that depends on the tagging response $r$, defined by
\begin{equation}\label{eq:pdf_bplus}
{\cal P}(r) = \left\{
\begin{array}{ll}
\etag   ( 1- \mistag ) & \mbox{$r$=``right tag decision''}  \\
\etag ~ \omega & \mbox{$r$=``wrong tag decision''}  \\
1-\etag & \mbox{$r$=``no tag decision''.}  \end{array}
\right. 
\end{equation}
The parameters $\omega$ and \etag~(defined in Eq.~\ref{eq:tageffdef})
are different for signal and background.
Fig.~\ref{fig:mass_bplus} shows the mass distribution of the selected and tagged events, together with the superimposed fit. 

%%%%%%%%%%%
\subsection{Analysis of the $\boldsymbol{\dstarmunu}$ channel}\label{sec:Dstmunu}
%%%%%%%%%%%

The \dstarmunu channel is selected by requiring that a muon and  
the decay $ \Dstarm \to \Dzb (\to K^+ \pi^-) \pi^-$
originate from a common vertex, displaced with respect to the $pp$ 
interaction point. 
The muon and \Dzb transverse momenta are required to be larger than  
0.8~\gevc and 1.8~\gevc respectively.
The selection criteria exploit the long \Bz and \Dzb lifetimes
by applying cuts on the impact parameters of the daughter tracks, 
on the pointing of the reconstructed \Bz momentum to the 
primary vertex, on the difference between the $z$ coordinate 
of the \Bz and \Dzb vertices, and on the \Dzb flight distance.
Additional cuts are applied on the muon and kaon
particle identification and on the quality of the fits of all tracks and 
vertices.
In case of multiple $B$ candidates per event the one 
with the smallest impact parameter significance with respect to the primary 
vertex is considered.
Only events triggered in the HLT by a single particle with large momentum, 
large transverse momentum and large $\rm IP$ are used.
In total, the sample consists of $\sim$482\,000 signal events.\\
Even though the final state is only partially reconstructed due to the missing 
neutrino, the contamination of background is small and 
the background to signal ratio $B/S$ is measured to be $\sim0.14$ in the 
signal mass region. 
The main sources of background are events containing a \Dzb originating from a 
$b$-hadron decay (referred to as \Dzb-from-$B$), %($5.5\pm0.1$\%); 
events with a $D^{*-}$ not from a $b$-hadron decay, %($3.6\pm0.1$\%);
decays of \Bp mesons to the same particles as the signal together 
with an additional pion (referred to as {\Bp}) %$3.3\pm 1.9$\%);  
and combinatorial background. %($1.3\pm0.1$\%). 
The different background sources can be disentangled from the signal by 
exploiting the different distributions of the observables $m$$=$$m_{K\pi}$, 
$\Delta m$$=$$m_{K\pi\pi}$$-$$m_{K\pi}$, the reconstructed \Bz decay time $t$ 
and the mixing state $q$.
The mixing state is determined by comparing the flavour of the reconstructed 
signal \Bd at decay time with the flavour indicated by the 
tagging decision (flavour at production time). 
For unmixed (mixed) events $q$$=$$+$$1$($-$$1$) while for untagged events $q$$=$$0$.
The decay time is calculated using the measured \Bz decay length, 
the reconstructed \Bz momentum and a correction for the missing neutrino
determined from simulation. It is  
parametrized as a function of the reconstructed \Bz invariant mass.

An extended unbinned maximum likelihood fit is performed by defining a PDF 
for the observables ($m,\Delta m, t, q$) as a product of one PDF for the masses 
and one for the $t$ and $q$ observables.
For the \Dzb and $D^{*-}$ mass peaks two double Gaussian distributions with common mean 
are used, while a parametric function motivated by available phase space 
is used to describe the $\Delta m$  
distributions of the \Dzb-from-$B$, and combinatorial background components. 
The decay time distribution of the signal consists of mixed, unmixed 
and untagged events, and is given by
\begin{equation}
{\cal P}^{\rm s}(t,q) \propto \left\{
\begin{array}{ll} 
\etag  ~ a(t)  \left\{ e^{-t/ \tau_{B^0}} \left[ 1+q(1-2\omega)  \cos(\dmd t) \right] \otimes R(t-t') \right\} & 
\mbox{ if $q=\pm1$} \\ 
(1-\etag)  a(t) \left\{ e^{-t/ \tau_{B^0}} \otimes R(t-t') \right\} & \mbox{ if $q=0$}, \\ 
\end{array}
\right. 
\label{eq:signalTime}
\end{equation}
where \dmd\ and $\tau_{B^0}$ are the \Bz--\Bzb mixing frequency and \Bz
lifetime. The decay time acceptance function is denoted by  $a(t)$ and  
$R(t-t')$ is the resolution model, both extracted from simulation. 
A double Gaussian distribution with common mean is used for the decay
time resolution model. 
In Eq.~\ref{eq:signalTime} the tagging parameters are assumed to be the 
same for $B$ and $\bar B$-mesons.

\begin{figure}[ht]
  	\begin{minipage}[b]{0.5\linewidth}
	\centering
	\begin{overpic}[width=0.95\textwidth]{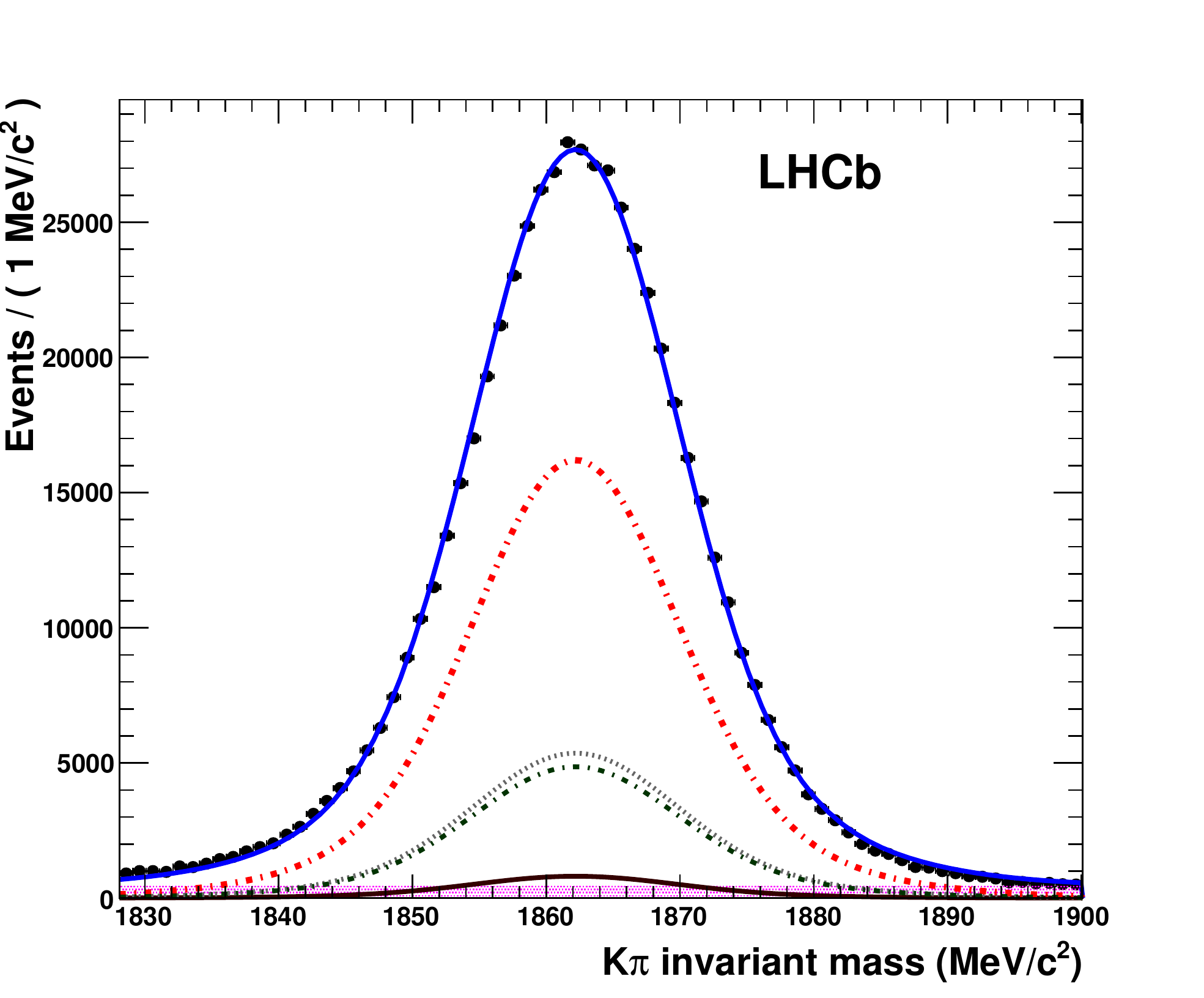}  
	\put (80,68) {\small{(a)}}		
	\end{overpic}
	\end{minipage}
  	\begin{minipage}[b]{0.5\linewidth}
	\centering
	\begin{overpic}[width=0.95\textwidth]{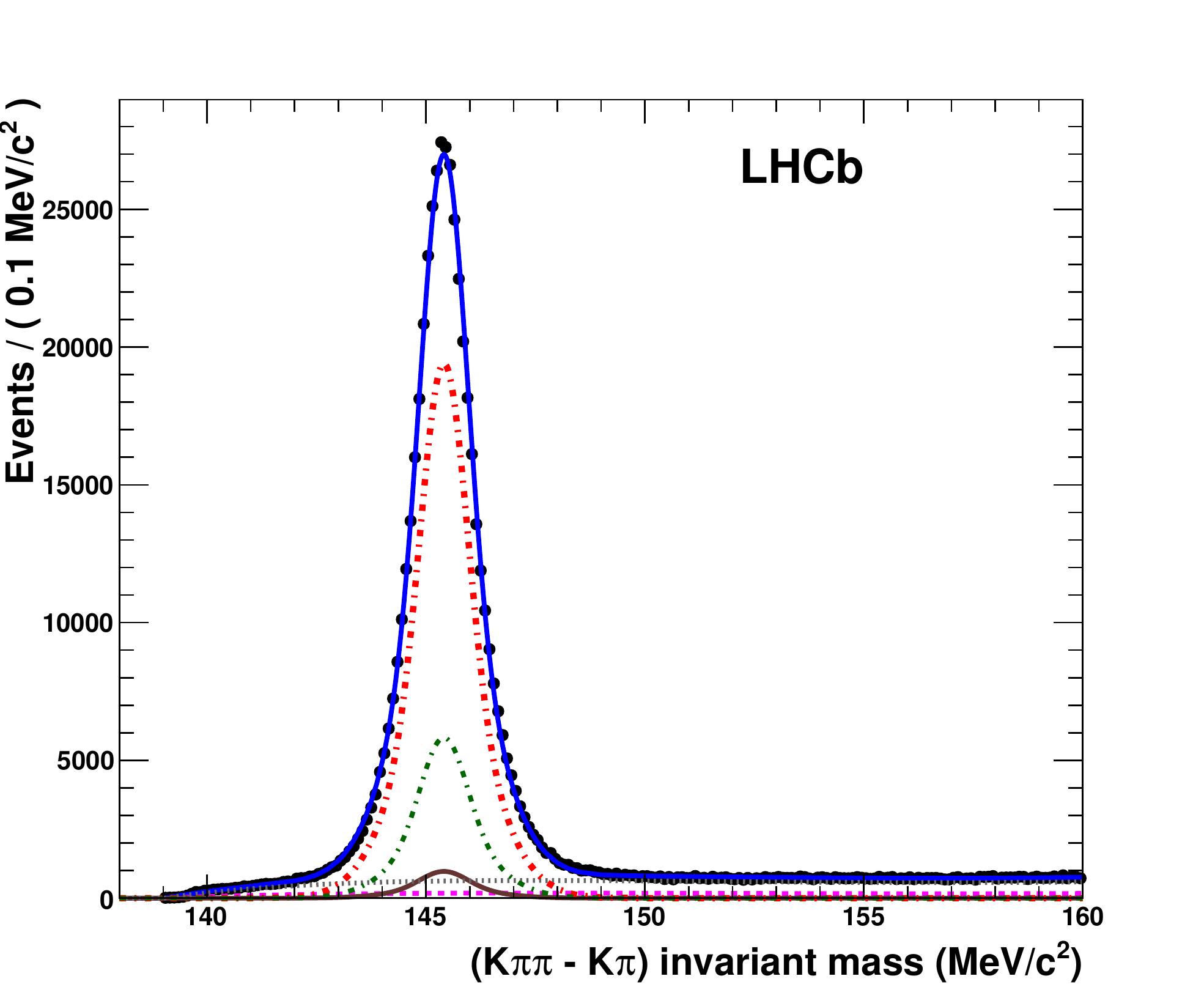}  
	\put (80,68) {\small{(b)}}		
	\end{overpic}
	\end{minipage}
  	\begin{minipage}[b]{0.5\linewidth}
	\centering
	\begin{overpic}[width=0.95\textwidth]{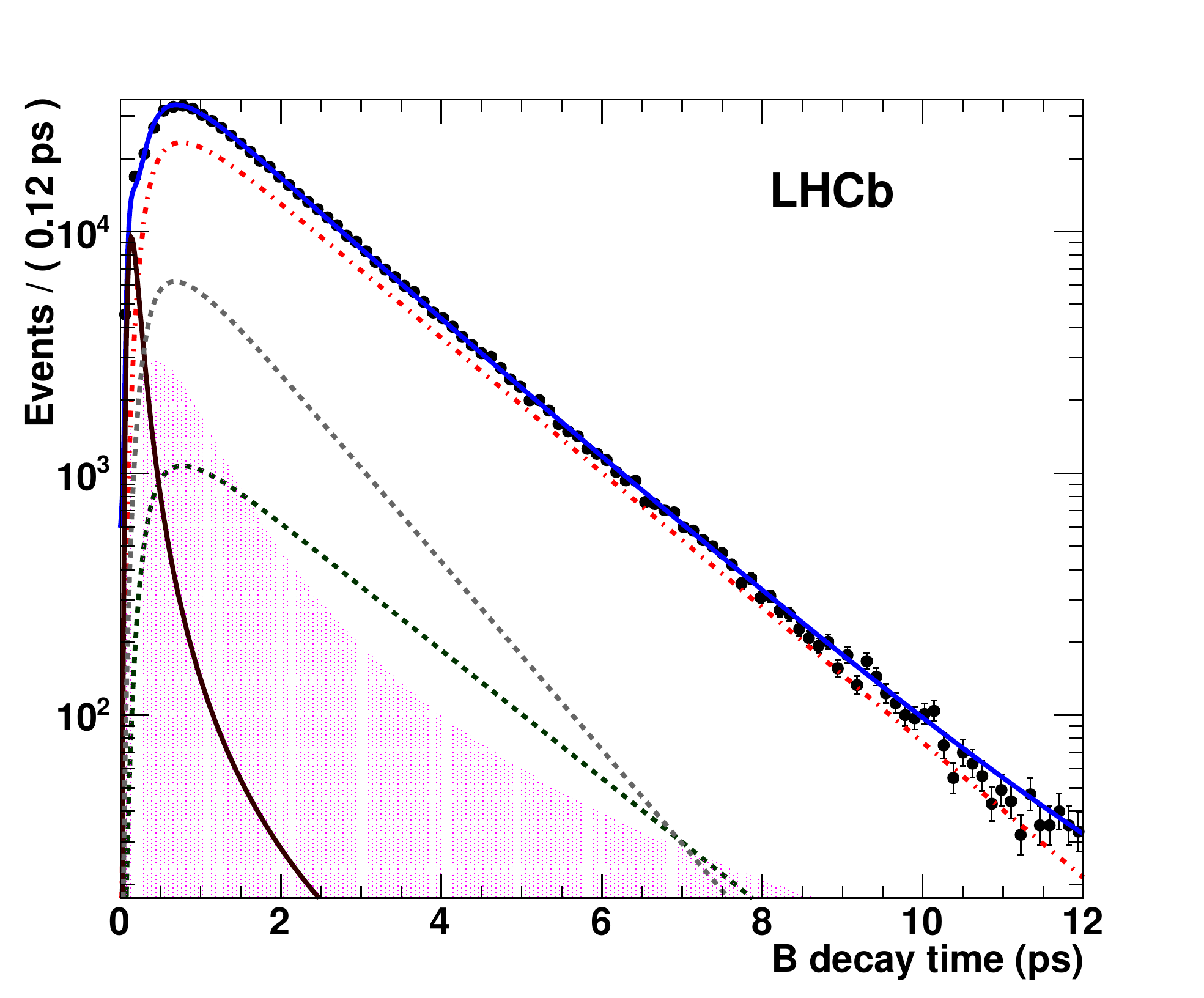} 
	\put (80,68) {\small{(c)}}	
	\end{overpic}
	\end{minipage}
     \caption{\small{
Distributions of (a) $K^+\pi^-$ invariant mass, 
(b) mass difference $m(K\pi\pi)$$-$$m(K\pi)$ and 
(c) decay time of the \dstarmunu events. 
Black points with errors are data, the blue curve is the fit result. 
The other lines represent signal (red dot-dashed), \Dzb-from-$B$ decay background (gray dashed), 
\Bu background (green short dashed), $D^*$ prompt background (magenta solid).
The combinatorial background is the magenta filled area.}}
    \label{fig:plots_dstarmunu}
\end{figure}

\begin{figure}[ht]
  \begin{center}
    \includegraphics[angle=0, width=0.5\textwidth]{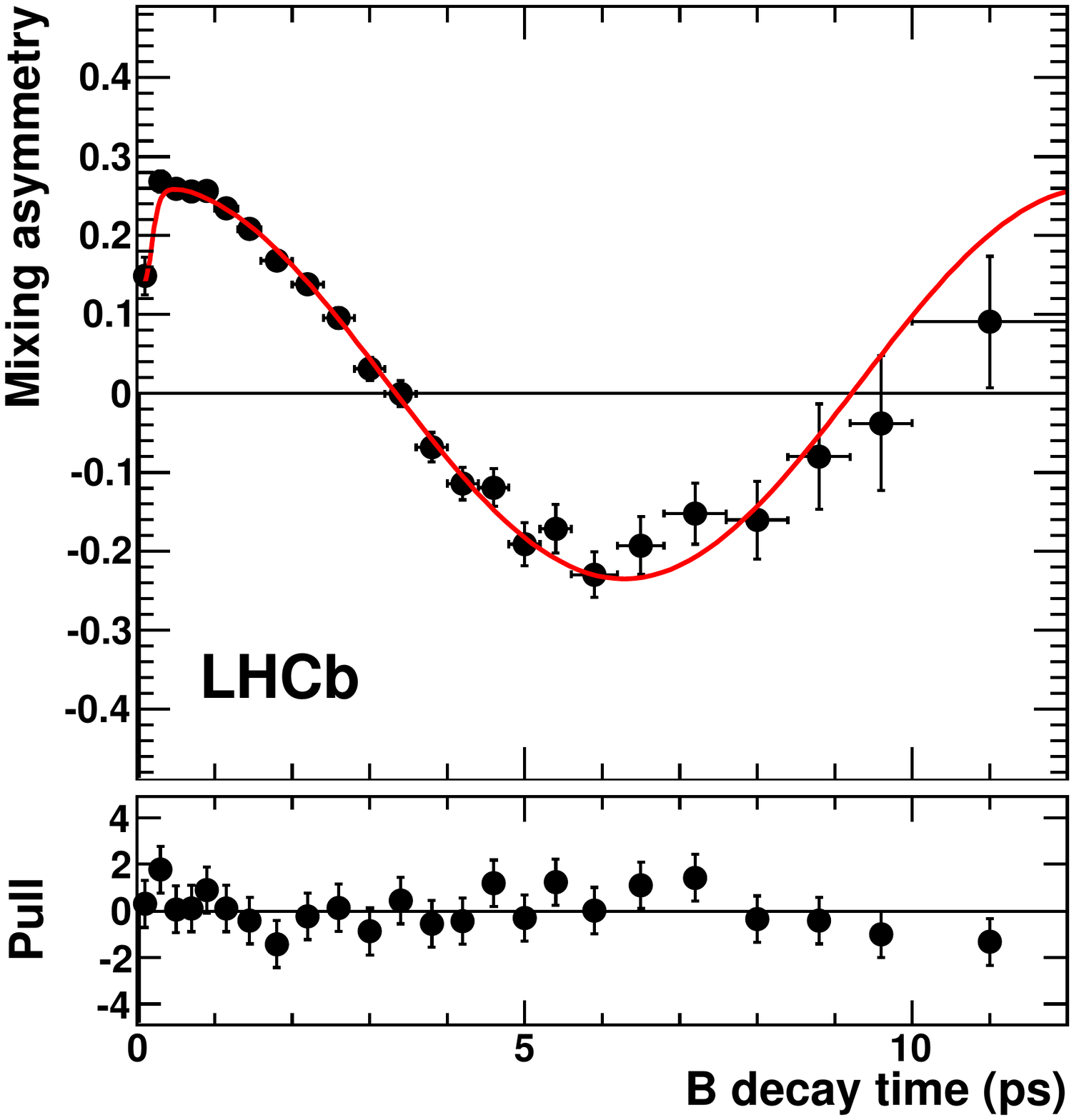}
     \end{center}
     \caption{\small{
     Raw mixing asymmetry of  \dstarmunu events  in the signal mass region  
when using the combination of all OS taggers. 
Black points are data and the red solid line is the result of the fit.
The lower plot shows the pulls of the residuals with respect to the fit.}}
    \label{fig:DMd_dstarmunu}
\end{figure}

The decay time distributions for the {\Bp} and  \Dzb-from-$B$ background 
components are taken as exponentials convolved by the resolution model  and 
multiplied by the same acceptance function as used for the signal.
For the prompt $D^*$ and combinatorial background, Landau 
distributions with independent parameters are used. The dependence on 
the mixing observable $q$ is the same as for the signal.
The tagging parameters \etag~ and \mistag~of the signal and of each 
background component are varied independently in the fit, except for 
the {\Bp} background where they are assumed to be equal to the parameters in the signal decay.
Figure~\ref{fig:plots_dstarmunu} shows the distributions of the mass and 
decay time observables used in the maximum likelihood fit.
The raw asymmetry is defined as
\begin{equation}
{\cal  A^{\rm raw}}(t) = 
\frac{N^{\rm unmix}(t)-N^{\rm mix}(t)}{N^{\rm unmix}(t)+N^{\rm mix}(t)}
\label{eq:mix-asymmetry1}
\end{equation}
where $N^{\rm mix}$ ($N^{\rm unmix}$) is the number of tagged events which 
have (not) oscillated at decay time $t$.
From Eq.~\ref{eq:signalTime} it follows that the asymmetry for signal 
is given by
 \begin{equation}
{\cal  A}(t) = (1-2\mistag )\cos (\dmd \, t) .
\label{eq:mix-asymmetry2}
\end{equation}
Figure~\ref{fig:DMd_dstarmunu} shows the raw asymmetry for the subset of events 
in the signal mass region that are tagged with the OS tagger combination.
At small decay times the asymmetry decreases due to the contribution
of background events,  ${\cal  A} \simeq 0$.
The value of \dmd\  was fixed to \dmd\ $=0.507$~\invps~\cite{PDG}. 
Letting the \dmd\ parameter vary in the fit gives consistent results. 


%%%%%%%%%%%
\subsection{Analysis of the $\boldsymbol{B^0 \to J/\psi K^{*0}}$ channel}\label{sec:bkstar}
%%%%%%%%%%%

The \bkstar channel is used to extract the mistag rate 
through a fit of the flavour oscillation of the \Bd mesons as a function of the decay time. The flavour of the \Bd meson at production time is determined from the tagging algorithms, while the flavour at the decay time is determined from
 the \Kstarz flavour, which is in turn defined by the kaon charge.

The \bkstar candidates are selected from \jpsi \to \mumu and 
\Kstarz \to $K^+\pi^-$ decays. 
The \jpsi mesons are selected by the same selection as used for the \bplus channel, described in Sect.~\ref{sec:bplus}.
The \Kstarz candidates are reconstructed from two good quality charged tracks 
identified as $K^{+}$ and $\pi^{-}$. 
The reconstructed \Kstarz meson is required to have a transverse momentum higher than 1\gevc,  a good quality vertex 
and an invariant mass within $\pm$\,70\mevcc of the nominal \Kstarz mass.
Combinations of \jpsi and \Kstarz candidates are accepted as \Bd candidates 
if they form a common vertex with good quality
and an invariant mass in the range $5100-5450$\mevcc. 
The \Bd transverse momentum is required to be higher than 2~\gevc.
The decay time and the invariant mass of the \Bd are extracted from a 
vertex fit with an identical procedure as for the \bplus channel, by applying
a constraint to the associated primary vertex, and a constraint to the \jpsi 
mass. 
In case of multiple $B$ candidates per event, only the candidate with the 
smallest $\chi^2$ of the vertex is kept. 

Only events that were triggered by the ``lifetime unbiased'' selection 
are kept. 
The \Bd candidates are required to have a decay time higher than 0.3\,ps to remove the large combinatorial background due to prompt \jpsi production. The sample contains $\sim 33\,000$ 
signal events.

The decay time distribution of signal events is parametrized as in 
Eq.~\ref{eq:signalTime}, without the acceptance correction.
The background contribution, with a background to signal ratio $B/S\sim0.29$, is due to misreconstructed $b$-hadron decays, where a dependence on the decay time is expected (labelled ``long-lived'' background). We distinguish two long-lived components.  
The first corresponds to events where one or more of the four tracks originate 
from a long-lived particle decay, but where the flavour of the reconstructed 
\Kstarz is not correlated with a true $b$-hadron.
Its decay time distribution is therefore modelled by a decreasing exponential. 
In the second long-lived background component, one of the tracks used to build 
the \Kstarz  originated from the primary vertex, hence 
the correlation between the \Kstarz and the $B$ flavour is partially lost. 
Its decay time distribution is more ``signal-like'', i.e. it is a decreasing 
exponential with an oscillation term, but with different mistag fraction and 
lifetime, left as free parameters in the fit.

\begin{figure}[ht]
  \begin{center}
    \includegraphics[angle=0, width=0.50\textwidth]{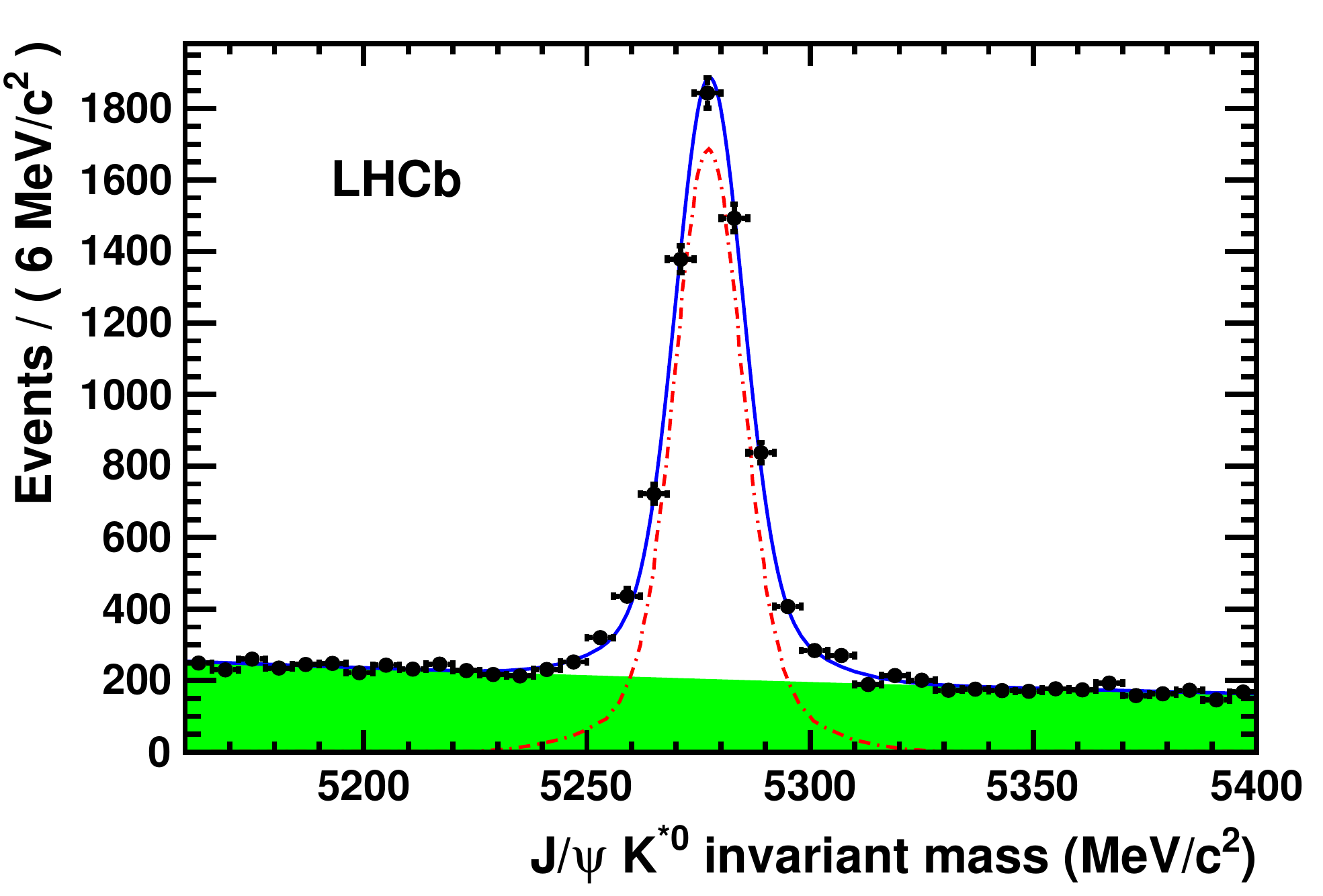}
     \end{center}
     \caption{\small{Mass distribution of OS tagged \bkstar events. 
Black points are data, the solid blue line, red dotted line and green 
area are the overall fit, the signal and the background components, 
respectively.}}
    \label{fig:mass_bkstar}
\end{figure}

The signal and background decay time distributions are convolved with the same resolution function, extracted from data.
The mass distributions, shown in Fig.~\ref{fig:mass_bkstar}, are described by a double Gaussian distribution peaking at the \Bz mass for the signal component, 
and by an exponential with the same exponent for both long-lived backgrounds.

\begin{figure}[ht]
  \begin{center}
    \includegraphics[angle=0, width=0.5\textwidth]{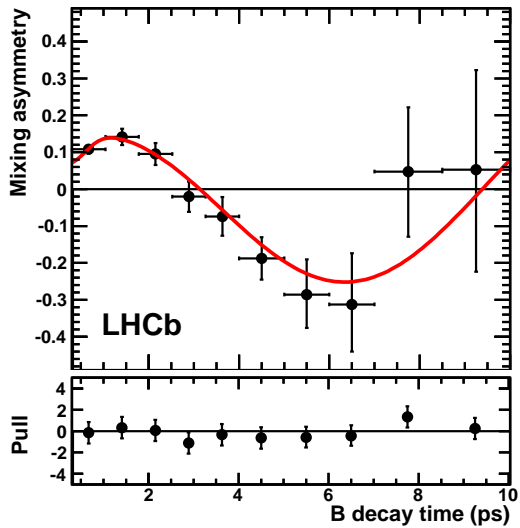}
     \end{center}
     \caption{\small{Raw mixing asymmetry of the \bkstar events  in the signal mass region, for all OS tagged events. 
Black points are data and the red solid line is the result of the fit.
The lower plot shows the pulls of the residuals with respect to the fit.
}}
    \label{fig:bkstar_asym}
\end{figure}

The OS mistag fraction is extracted from a fit to all tagged data,
with the values for the \Bd lifetime and \dmd\ fixed to the world 
average~\cite{PDG}.
Figure~\ref{fig:bkstar_asym} shows the time-dependent mixing asymmetry in the 
signal mass region, obtained using the information of the OS tag decision.
Letting the \dmd\ parameter vary in the fit gives consistent results. 

%%%%%%%%%%%
\section{Calibration of the mistag probability on data}
\label{sec:tagging_calibration}
%%%%%%%%%%%
For each individual  tagger and for the combination of taggers, the calculated 
mistag probability ($\eta$) is obtained on an event-by-event basis 
from the neural network output.
The values are calibrated in a fit using the measured mistag fraction 
(\mistag) from the self-tagged control channel \bplus. 
A linear dependence between the measured and the calculated mistag probability 
for signal events is used, as suggested by the data distribution,
\begin{equation}\label{eq:calibration}
\mistag(\eta) = p_0 + p_1  (\eta - \langle\eta\rangle)\; ,
\end{equation}
where $p_0$ and $p_1$ are parameters of the fit and $\langle\eta\rangle$ 
is the mean 
calculated mistag probability. This parametrization is chosen to minimize the
correlation between the two parameters.
Deviations from $p_0 = \langle\eta\rangle$ and $p_1=1$ would indicate that 
the calculated mistag probability should be corrected. 

In order to extract the $p_0$ and $p_1$  calibration parameters, an unbinned 
maximum likelihood fit to the mass, tagging decision and mistag probability 
$\eta$ observable is performed.
The fit parametrization takes  into account the probability density function 
of $\eta$, $\cal P(\eta)$, that is extracted from data for signal and 
background separately, using events in different mass regions. 
For example, the PDF for signal events from Eq.~\ref{eq:pdf_bplus} 
then becomes
\begin{equation}
{\cal P^{\rm s}}(r,\eta) = 
 \left\{
\begin{array}{ll}
\etag  \left( 1- \mistag(\eta) \right) \cal P^{\rm s}(\eta)  & \mbox{$r$=``right tag decision''}  \\
\etag ~ \omega(\eta) \cal P^{\rm s}(\eta)  & \mbox{$r$=``wrong tag decision''}  \\
1-\etag & \mbox{$r$=``no tag decision''.}  \end{array}
\right. 
\end{equation}
\noindent The measured mistag fraction of the background is assumed to be 
independent from the calculated mistag probability, as confirmed by the 
distribution of background events.

The calibration is performed on part of the data sample in a two-step 
procedure.  
Each tagger is first calibrated individually. The results show that, for each 
single tagger, only a minor adjustment of $p_0$ with respect to the starting 
calibration of the neural network, performed on simulated events, is required.
In particular, the largest correction is 
$p_0 -$~$\langle\eta\rangle =$ 0.033$\pm$0.005 in the case of the vertex 
charge tagger, while the deviations from unity of the $p_1$ parameter are 
about 10\%, similar to the size of the corresponding statistical errors.
In a second step the calibrated mistag probabilities are combined and finally 
the combined mistag probability 
is calibrated. This last step is necessary to correct for the small 
underestimation ($p_0 - \langle \eta \rangle =$ 0.022$\pm$0.003)
of the combined mistag probability due to the correlation 
among taggers neglected in the combination procedure.
The calibrated mistag is referred to as $\eta_c$ in the following.

\begin{figure}[h]
  \begin{center}
    \includegraphics[angle=0, width=0.50\textwidth]{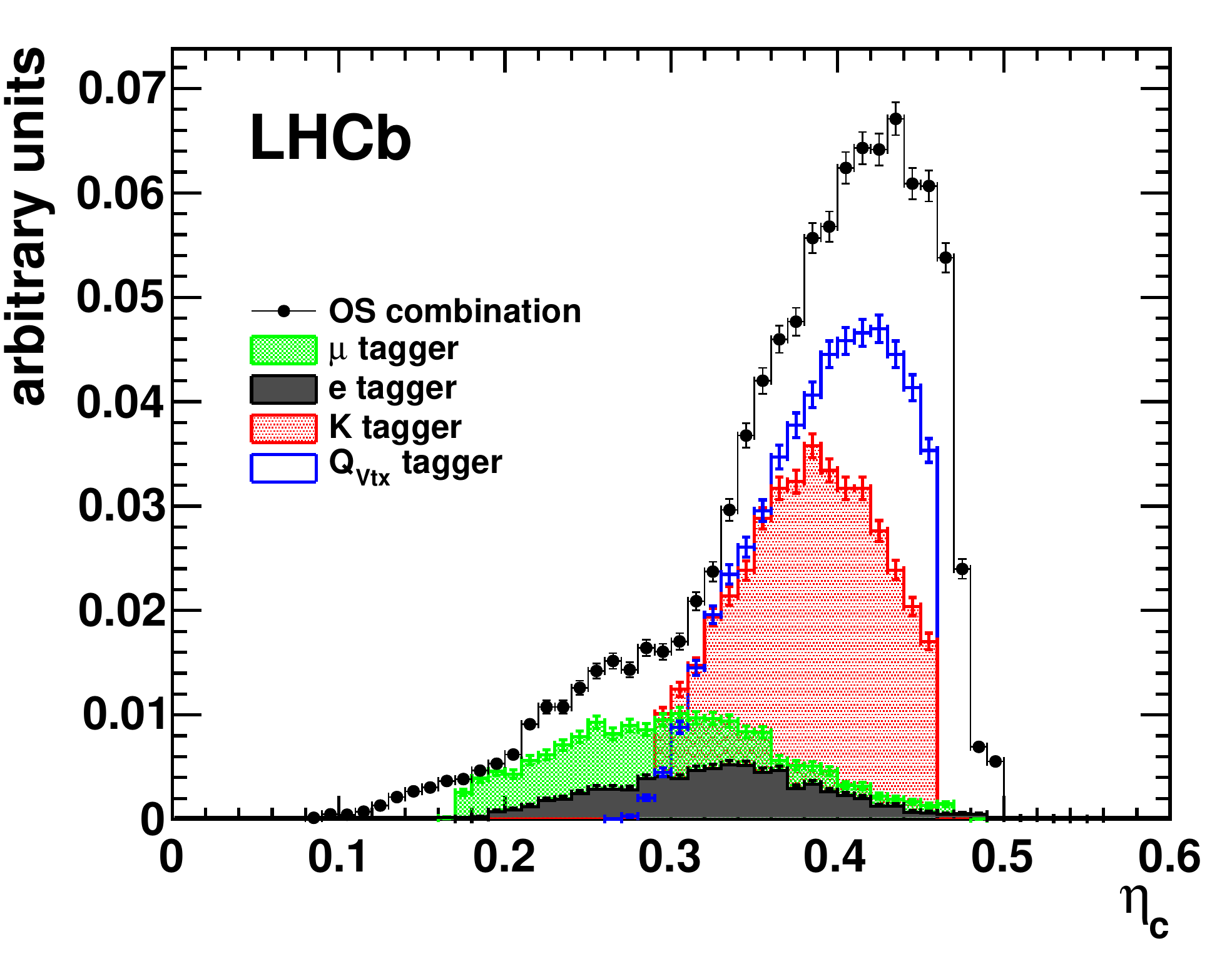}    
\caption{\small{Distribution of the calibrated mistag probability for the 
single OS taggers and their combination for \bplus  
events selected in a $\pm24$ \mevcc mass window around the \Bu mass. 
}}
\label{fig:mistagDistributions} 
  \end{center}
\end{figure}

Figure~\ref{fig:mistagDistributions} shows the distribution of the mistag 
probability for each tagger and for their combination, as obtained for 
\bplus events selected in a $\pm24$~\mevcc mass window 
around the \Bu mass.
%%%%%%%%%%%
\section{Tagging performance}\label{sec:results}
%%%%%%%%%%%

The tagging performances of the single taggers and of the
OS  combination measured after the calibration 
of the mistag probability are shown in Tables~\ref{tab:bplus_cat},
\ref{tab:bkstar_cat} and~\ref{tab:dstarmunu_cat} for the 
\bplus,~\bkstar and \dstarmunu channels, respectively.

The performance of the OS combination is evaluated in different ways.
First the average performance of the OS combination is calculated,
giving the same weight  to each event.
In this case, the  best tagging power is obtained by rejecting the events 
with a poor predicted mistag probability $\eta_c$ (larger than $0.42$), 
despite a lower \etag.
Additionally, to better exploit the tagging information, the tagging performance 
is determined on independent samples obtained by binning the data in 
bins of $\eta_c$.
The fits described in the previous sections are repeated for each sub-sample, 
after which the tagging performances are determined. 
As the samples are independent, the tagging efficiencies and the tagging powers 
are summed and subsequently the effective mistag is extracted.
The total tagging power increases by about 30\% with respect to the average 
value, as shown in the last line of 
Tables~\ref{tab:bplus_cat}-\ref{tab:dstarmunu_cat}.

The measured tagging performance is similar among the three channels.  
The differences between the \bplus and \bkstar results are large in absolute 
values, but still compatible 
given the large statistical uncertainties of the \bkstar results. 
Differences between the tagging efficiency in the \dstarmunu and the
$B \to J/\psi X$ channels were shown in previous MC studies to be related to 
the different $B$ momentum spectra and to different contributions 
to the trigger decision~\cite{lhcb-2007-058}.
\begin{table} [ht]
\caption{\small{Tagging performance in the \bplus channel. Uncertainties
are statistical only.}}
\begin{center}
\begin{tabular} { c | c | c | c  }
\hline
Taggers &  \etag  [\%] & $\omega$   [\%] & $\etag (1-2\omega)^2$  [\%]\\
\hline
$\mu$ & \,\, 4.8$\pm$0.1 & 29.9$\pm$0.7 & 0.77$\pm$0.07 \\
$e$         & \,\,  2.2$\pm$0.1 & 33.2$\pm$1.1 & 0.25$\pm$0.04 \\
$K$         & 11.6$\pm$0.1 & 38.3$\pm$0.5 & 0.63$\pm$0.06 \\
$Q_{\mathrm{vtx}}$ & 15.1$\pm$0.1 & 40.0$\pm$0.4 & 0.60$\pm$0.06\\
\hline
OS average ($\eta_c<$0.42)  &17.8$\pm$0.1 &34.6$\pm$0.4 &1.69$\pm$0.10 \\
\hline
OS sum of $\eta_c$ bins & 27.3$\pm$0.2 & 36.2$\pm$0.5 & 2.07$\pm$0.11 \\
\hline
\end{tabular}
\end{center}
\label{tab:bplus_cat}
\end{table}
\begin{table} [ht]
\caption{\small{Tagging performance in the \bkstar channel. 
Uncertainties are statistical only.}}
\begin{center}
\begin{tabular} { c | c | c | c }
\hline
Taggers &  \etag  [\%] & $\omega$   [\%] & $\etag (1-2\omega)^2$  [\%]\\
\hline
$\mu$ & \,\, 4.8$\pm$0.1 & 34.3$\pm$1.9 & 0.48$\pm$0.12 \\
$e$         & \,\, 2.2$\pm$0.1 & 32.4$\pm$2.8 & 0.27$\pm$0.10 \\
$K$         & 11.4$\pm$0.2 & 39.6$\pm$1.2 & 0.49$\pm$0.13 \\
$Q_{\mathrm{vtx}}$ & 14.9$\pm$0.2 & 41.7$\pm$1.1 & 0.41$\pm$0.11 \\
\hline
OS average ($\eta_c<$0.42) &17.9$\pm$0.2&36.8$\pm$1.0&1.24$\pm$0.20 \\
\hline
OS sum of $\eta_c$ bins  &27.1$\pm$0.3 & 38.0$\pm$0.9 & 1.57$\pm$0.22 \\ %
\hline
\end{tabular}
\end{center}
\label{tab:bkstar_cat}
\end{table}
\begin{table} [ht]
\caption{\small{Tagging performance in the 
$B^0 \to D^{*-} \mu^+ \nu_\mu$ channel. Uncertainties 
are statistical only.}}
\begin{center}
\begin{tabular} {c | c | c | c }
\hline
Taggers &  \etag  [\%] & $\omega$   [\%] & $\etag (1-2\omega)^2$  [\%]\\
\hline
$\mu$	 &  \,\,6.08$\pm$0.04 & 33.3$\pm$0.4 & 0.68$\pm$0.04\\	
e &  \,\,2.49$\pm$0.02& 34.3$\pm$0.7 &0.25$\pm$0.02 \\	
K & 13.36$\pm$0.05 &38.3$\pm$0.3 & 0.74$\pm$0.04\\
$Q_{\mathrm{vtx}}$ & 16.53$\pm$0.06 & 41.5$\pm$0.3 & 0.48$\pm$0.03 \\
\hline
OS average ($\eta_c<$0.42)  & 20.56$\pm$0.06& 36.1$\pm$0.3 & 1.58$\pm$0.06\\
\hline
OS sum of $\eta_c$ bins & 30.48$\pm$0.08  &37.0$\pm$0.3  & 2.06$\pm$0.06\\
\hline
\end{tabular}
\label{tab:dstarmunu_cat}
\end{center}
\end{table}

\section{Systematic uncertainties}
%=======================================================================
The systematic uncertainties on the calibration parameters $p_0$ and $p_1$ 
are studied by repeating the calibration procedure on \bplus events for 
different conditions.
The difference is evaluated between the value of the fitted parameter 
and the reference value, and 
is reported in the first row of Table~\ref{tab:OScal}.
Several checks are performed of which the most relevant are
reported in Table~\ref{table:tagsyst-body} and are described below:
%=======================================================================
\begin{table}[h]
\caption{\small{Fit values and correlations of the OS combined mistag 
calibration parameters measured in the \bplus, \bkstar and  \dstarmunu
channels. The uncertainties are statistical only.}}
\begin{center}
\begin{tabular} { c | c | c | c | c | c}
\hline
 Channel  &$p_0$& $p_1$ & $\langle\eta_c\rangle$ &  $p_0 - p_1 \langle\eta_c\rangle$ &$\rho(p_0,p_1)$\\
\hline 
\bplus     & $0.384 \pm 0.003 $ & $1.037 \pm 0.038$ & $0.379$ & $-0.009\pm 0.014$ & $0.14$ \\ 
\bkstar    & $0.399 \pm 0.008 $ & $1.016 \pm 0.102$ & $0.378$ & $\;\;\;0.015\pm0.039$  & $0.05$ \\ 
\dstarmunu & $0.395 \pm0.002$ & $1.022 \pm 0.026$ & $0.375$ &$\;\;\;0.008\pm0.010$  & $0.14$ \\ 
    \hline
  \end{tabular}
\end{center}
\label{tab:OScal}
\end{table}
% TTTTTTTTTTTTTTTTTTTTTTTTTTTTTTT
\begin{table}[h!]
\caption{\small{Systematic uncertainties on the calibration parameters 
$p_0$ and  $p_1$ obtained with \bplus events.}}
\begin{center}
\begin{tabular}{c|c|c|c} \hline
Systematic effect   &  $\delta p_0$ & $\delta p_1$ &  $\delta(p_0 - p_1\langle\eta_c\rangle)$\\ \hline
Run period & $\pm0.005$ & $\pm0.003$ & $\pm0.004$ \\ 
$B$-flavour           & $\pm0.008$ & $\pm0.067$ & $\pm0.020$ \\
Fit model assumptions ${\cal P}(\eta)$ & $<\pm0.001$ & $\pm0.005$ & $\pm0.002$ \\
\hline
Total  & $\pm0.009$& $\pm0.07$ & $\pm0.02$ \\ \hline
\end{tabular}
%}
\label{table:tagsyst-body}
\end{center}
\end{table}
% TTTTTTTTTTTTTTTTTTTTTTTTTTTTTTT
\begin{itemize}
\item The data sample is split according to the run periods and to the 
magnet polarity, in order
 to check whether possible asymmetries of the detector efficiency, or of the 
alignment accuracy, or variations in the data-taking conditions
introduce a difference in the tagging calibration.
\item The data sample is split according to the signal flavour,
 as determined by the reconstructed final state. 
In fact, the calibration of the mistag probability for different \B  
flavours might be different due to the different 
particle/antiparticle interaction with matter or possible detector asymmetries.
In this case a systematic uncertainty has to be considered, unless the 
difference is explicitly taken into account when fitting 
for \CP\ asymmetries.
\item The distribution of the mistag probability in the fit model, 
${\cal P}(\eta)$, is varied either by assuming the signal 
and background distributions to be equal or by swapping them. 
In this way possible uncertainties related to the fit model are considered.
\end{itemize}
In addition, the stability of the calibration parameters is verified
for different bins of transverse momentum of the signal $B$.

The  largest systematic uncertainty  in Table~\ref{table:tagsyst-body}
originates from the dependence on the signal flavour.
As a cross check this dependence is also  measured with \dstarmunu events,
repeating the calibration after splitting the sample according to the 
signal decay flavour. The differences in this case 
are $\delta p_0=\pm0.009$ and $\delta p_1=\pm0.009$, 
where the latter is smaller than in the \bplus channel. 
Both for the run period dependence
and for the signal flavour the variations of
$\delta p_0$ and $\delta p_1$ are not statistically significant.
However, as a conservative estimate of the total systematic uncertainty
on the calibration parameters, all the contributions 
in Table~\ref{table:tagsyst-body} are summed in quadrature.
The tagging efficiencies do not depend on the initial flavour of the signal 
$B$. In the case of the \bplus channel the values are $(27.4 \pm 0.2)$\% for 
the $B^+$ and $ (27.1 \pm 0.2) $\% for the $B^-$.

\section{Comparison of decay channels}
%=======================================================================
The dependence of the calibration of the OS mistag probability
on the decay channel is studied.
The values of $p_0$, $p_1$ and $\langle \eta_c \rangle$ 
measured on the whole data sample for all the three channels 
separately, are shown in Table~\ref{tab:OScal}. 
The parameters $p_1$ are compatible with 1,
within the statistical uncertainty. The differences
$p_0 - p_1 \langle \eta_c \rangle$, shown in the fifth column,
are compatible with zero, as expected.
In the last column the correlation coefficients are shown.

\begin{figure}
	\begin{minipage}[b]{0.5\linewidth}
	\centering
	\begin{overpic}[width=0.75\textwidth]{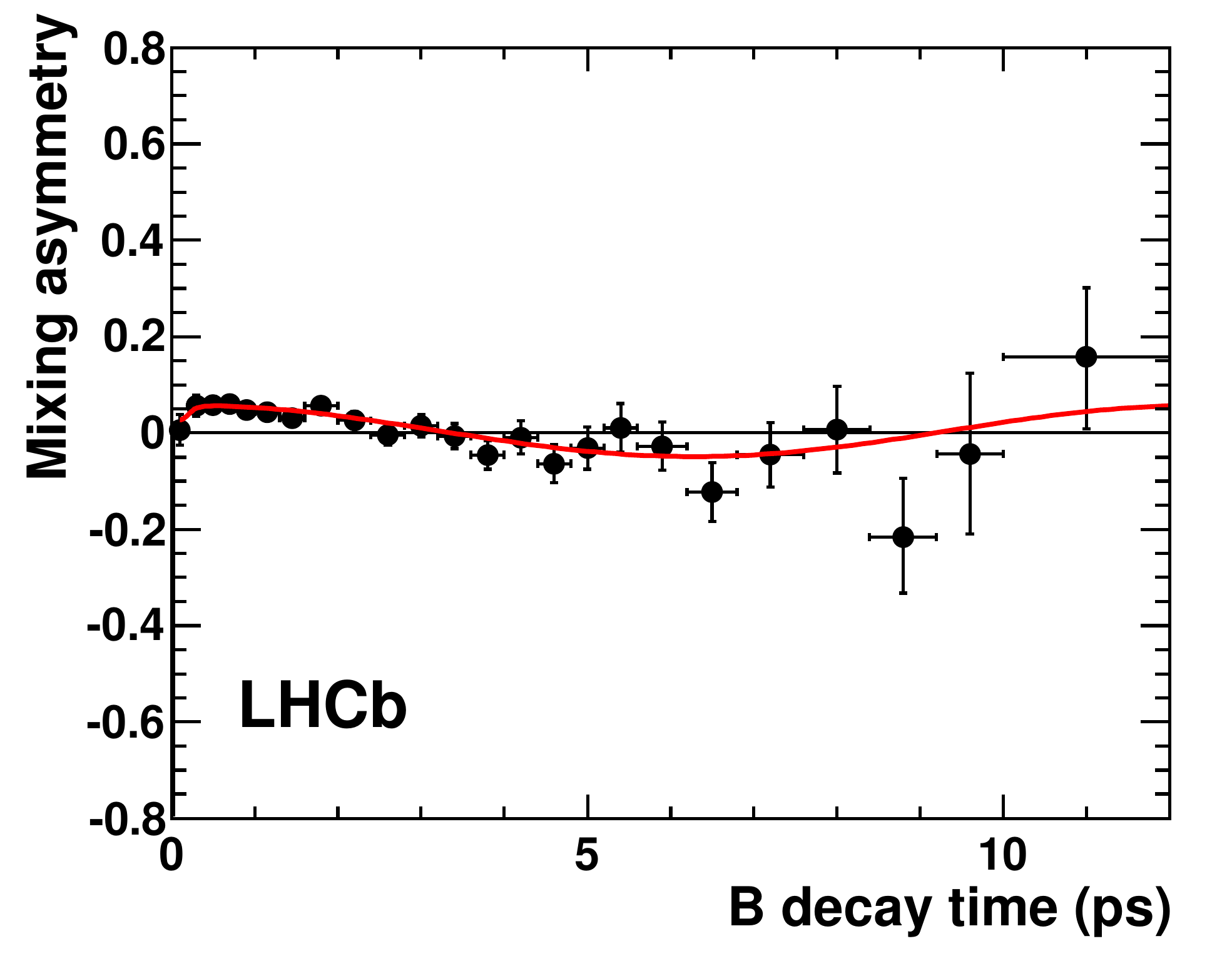}
	\put (35,68) {\small{$0.43 \leq \eta_c < 0.50$}}		
	\end{overpic}
	\end{minipage}
	\begin{minipage}[b]{0.5\linewidth}
	\centering
	\begin{overpic}[width=0.75\textwidth]{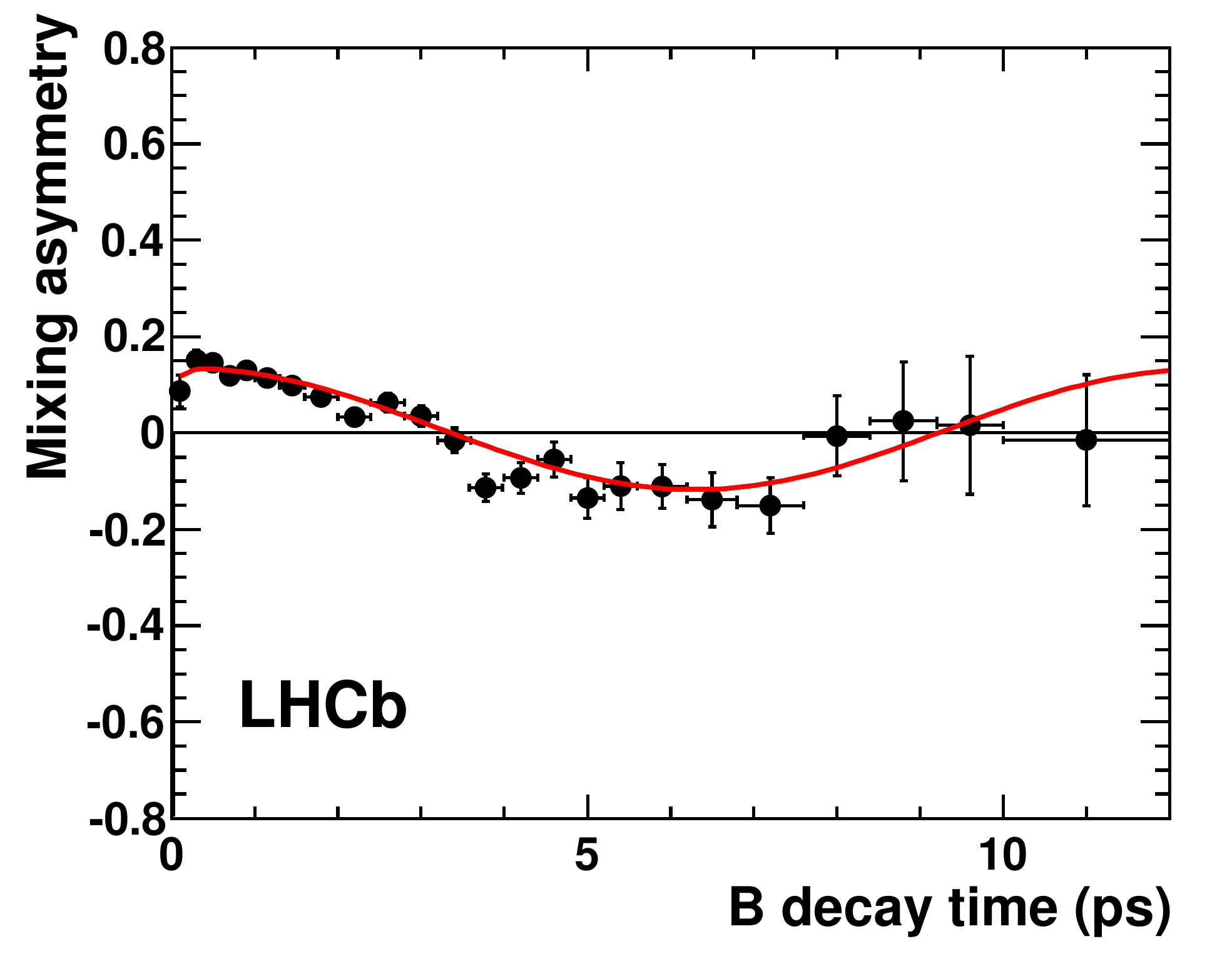}
	\put (35,68) {\small{$0.38 \leq \eta_c < 0.43$}}		
	\end{overpic}	
	\end{minipage}
	\begin{minipage}[b]{0.5\linewidth}
	\centering
	\begin{overpic}[width=0.75\textwidth]{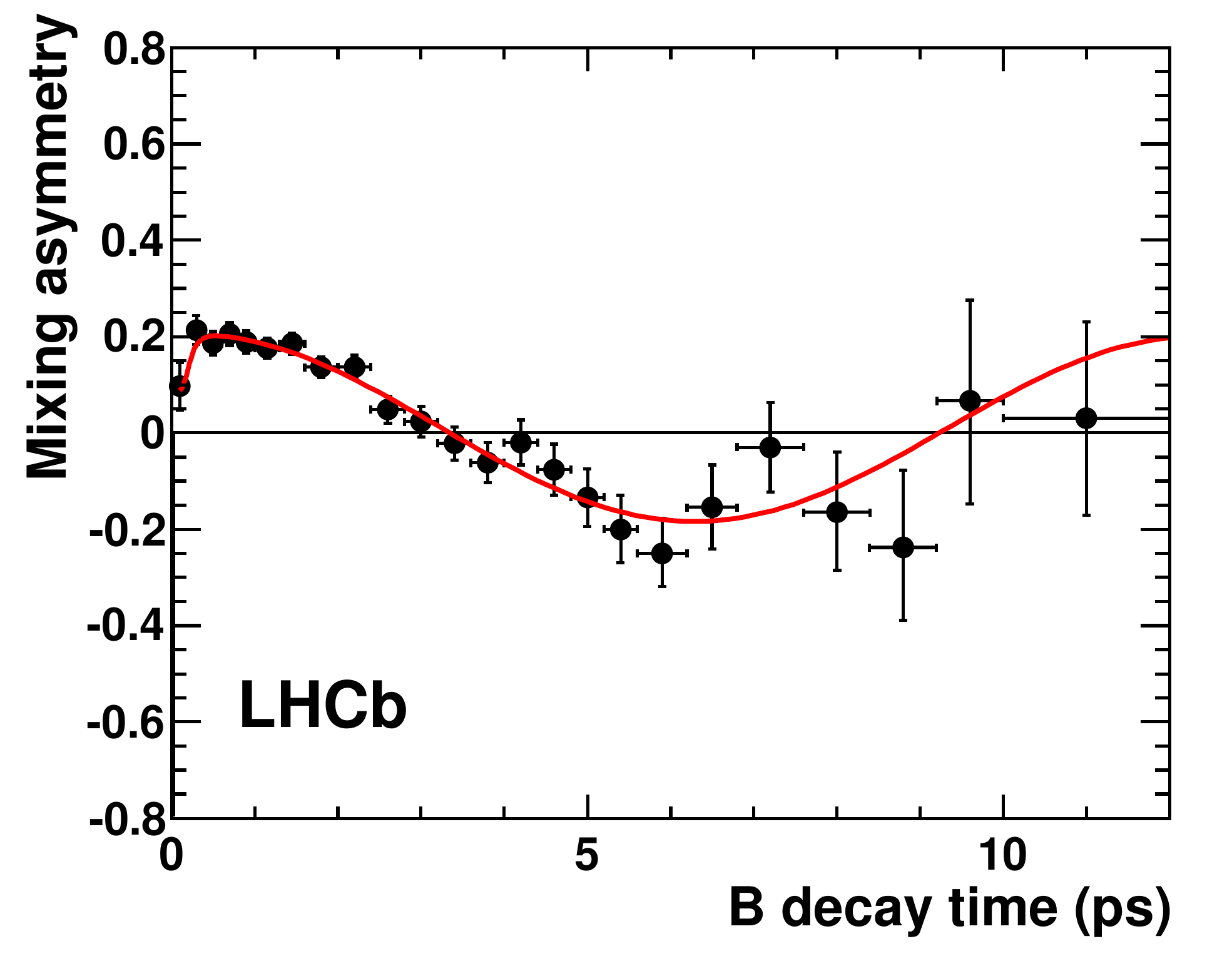}
	\put (35,68) {\small{$0.35 \leq \eta_c < 0.38$}}		
	\end{overpic}	
	\end{minipage}
	\begin{minipage}[b]{0.5\linewidth}
	\centering
	\begin{overpic}[width=0.75\textwidth]{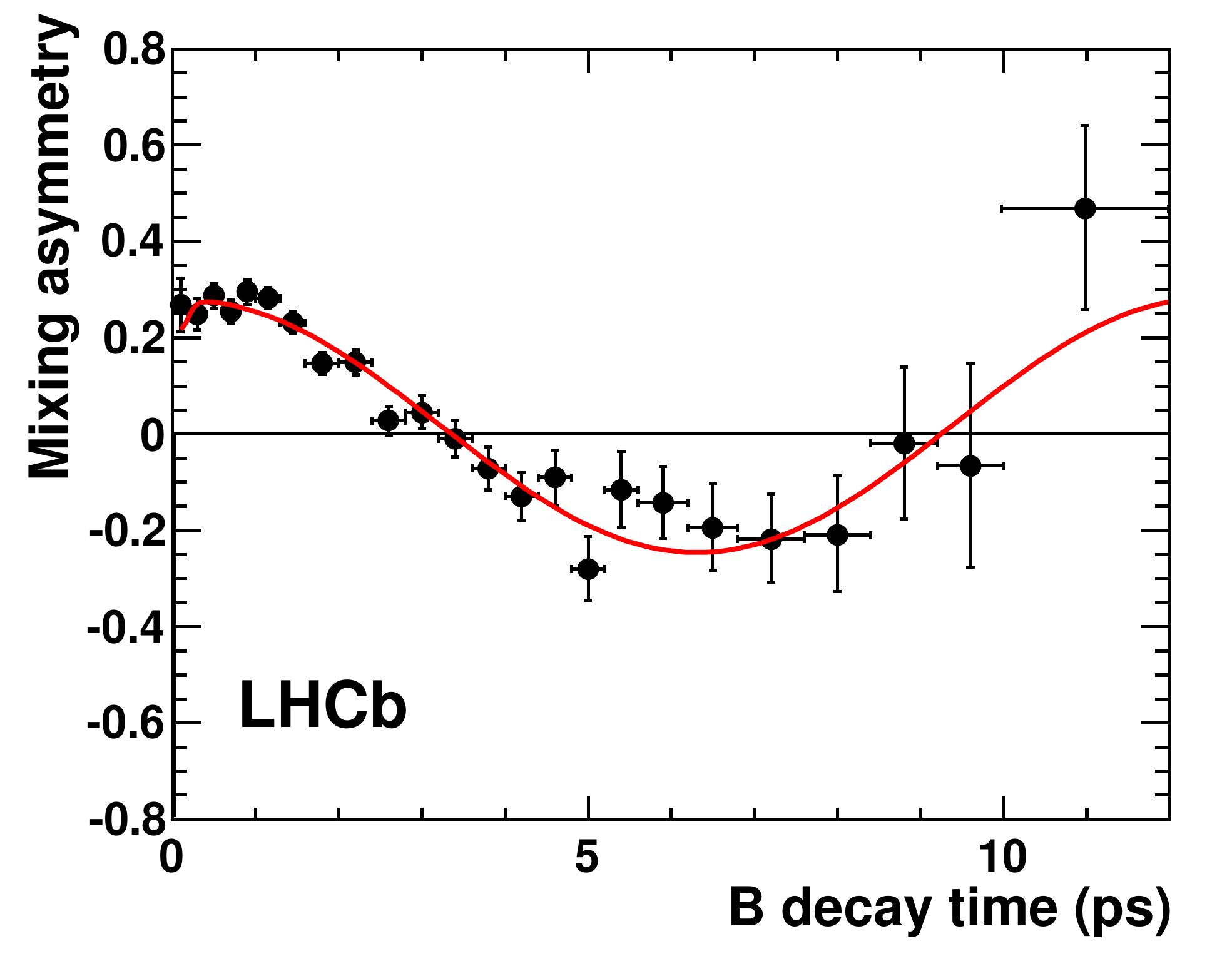}
	\put (35,68) {\small{$0.31 \leq \eta_c < 0.35$}}		
	\end{overpic}	
	\end{minipage}
	\begin{minipage}[b]{0.5\linewidth}
	\centering
	\begin{overpic}[width=0.75\textwidth]{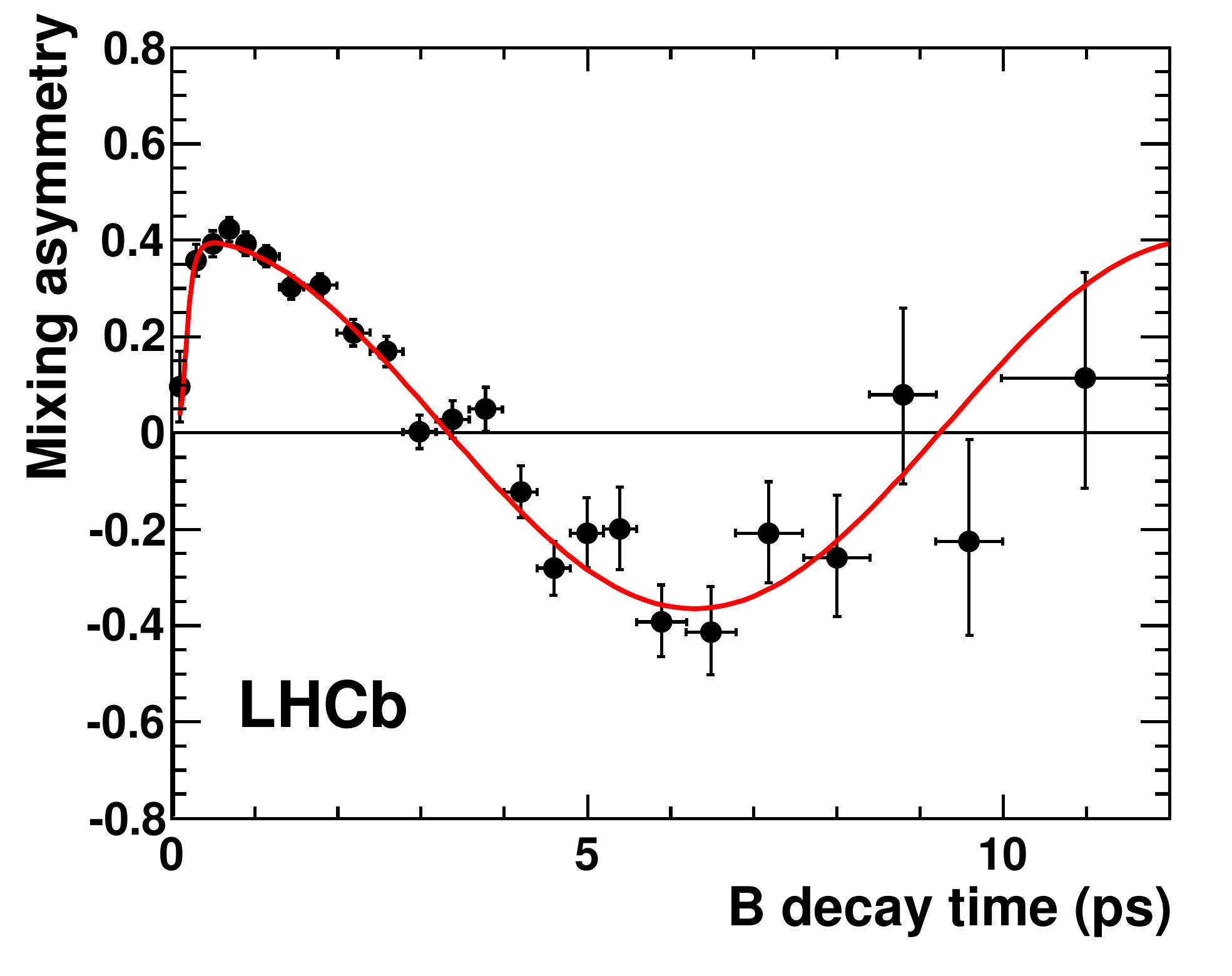}
	\put (35,68) {\small{$0.24 \leq \eta_c < 0.31$}}		
	\end{overpic}	
	\end{minipage}
	\begin{minipage}[b]{0.5\linewidth}
	\centering
	\begin{overpic}[width=0.75\textwidth]{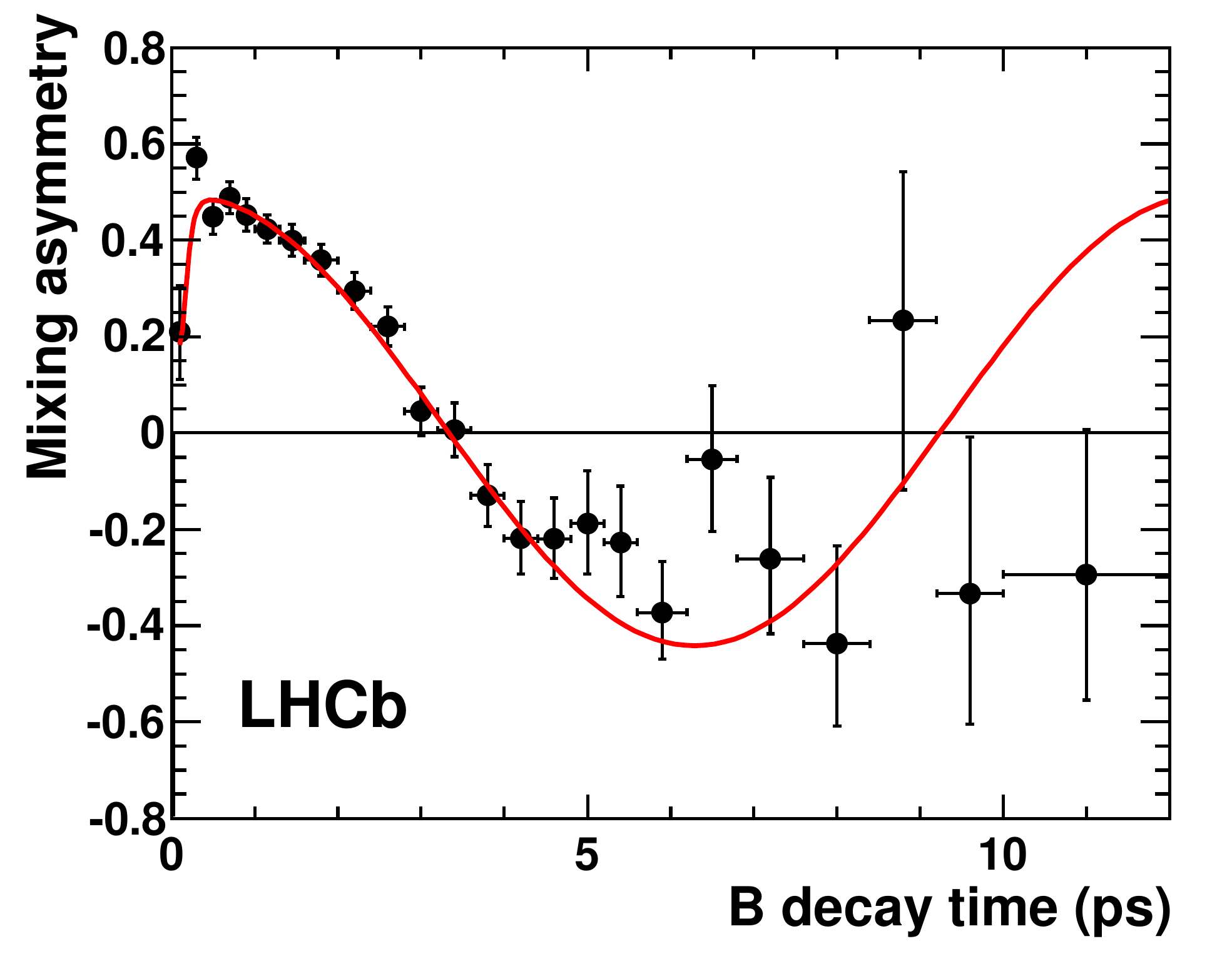}
	\put (35,68) {\small{$0.17 \leq \eta_c < 0.24$}}		
	\end{overpic}	
	\end{minipage}
	\begin{minipage}[b]{0.5\linewidth}
	\centering
	\begin{overpic}[width=0.75\textwidth]{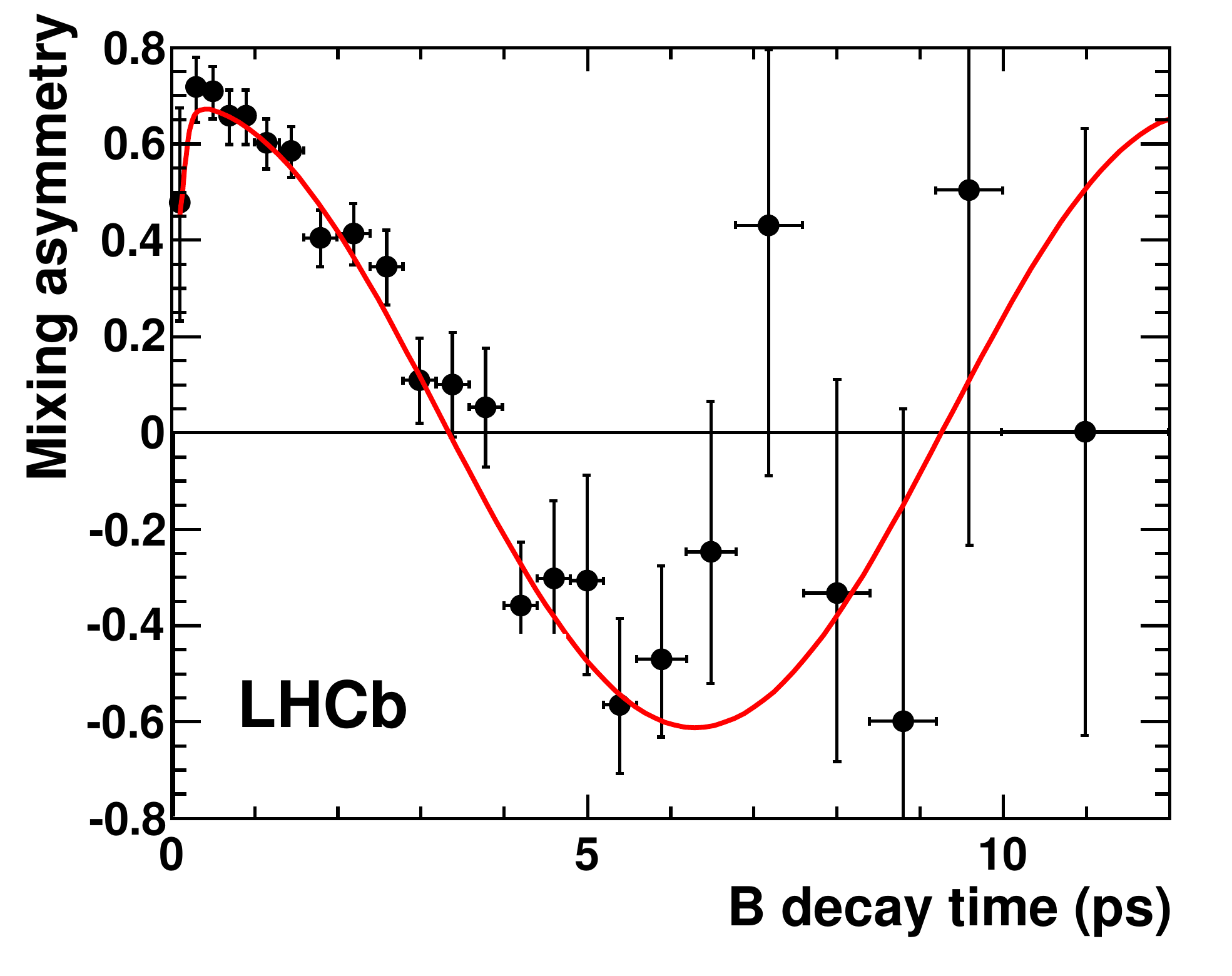}
	\put (35,68) {\small{$\eta_c < 0.17$}}		
	\end{overpic}	
	\end{minipage}
\caption{\small{Raw mixing asymmetry as a function of \B decay time in 
\dstarmunu events, in the signal mass region, using the OS tagger.
Events are split into seven samples of decreasing mistag probability $\eta_c$.
}}
\label{fig:dstarmu_oscillation}
\end{figure}

\begin{figure}[ht]
  \begin{center}
     \includegraphics[angle=0, width=0.45\textwidth]{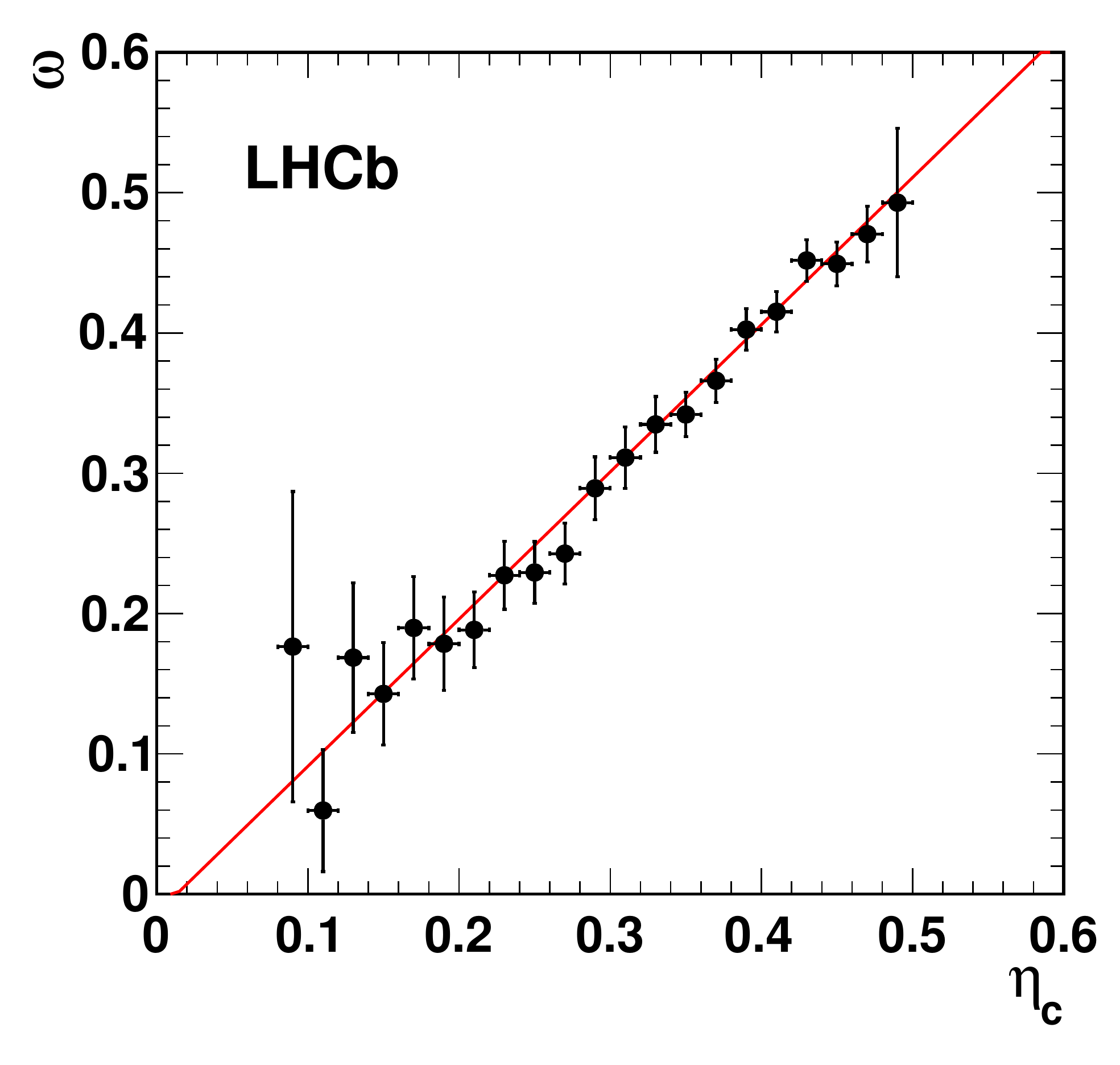}
     \includegraphics[angle=0, width=0.45\textwidth]{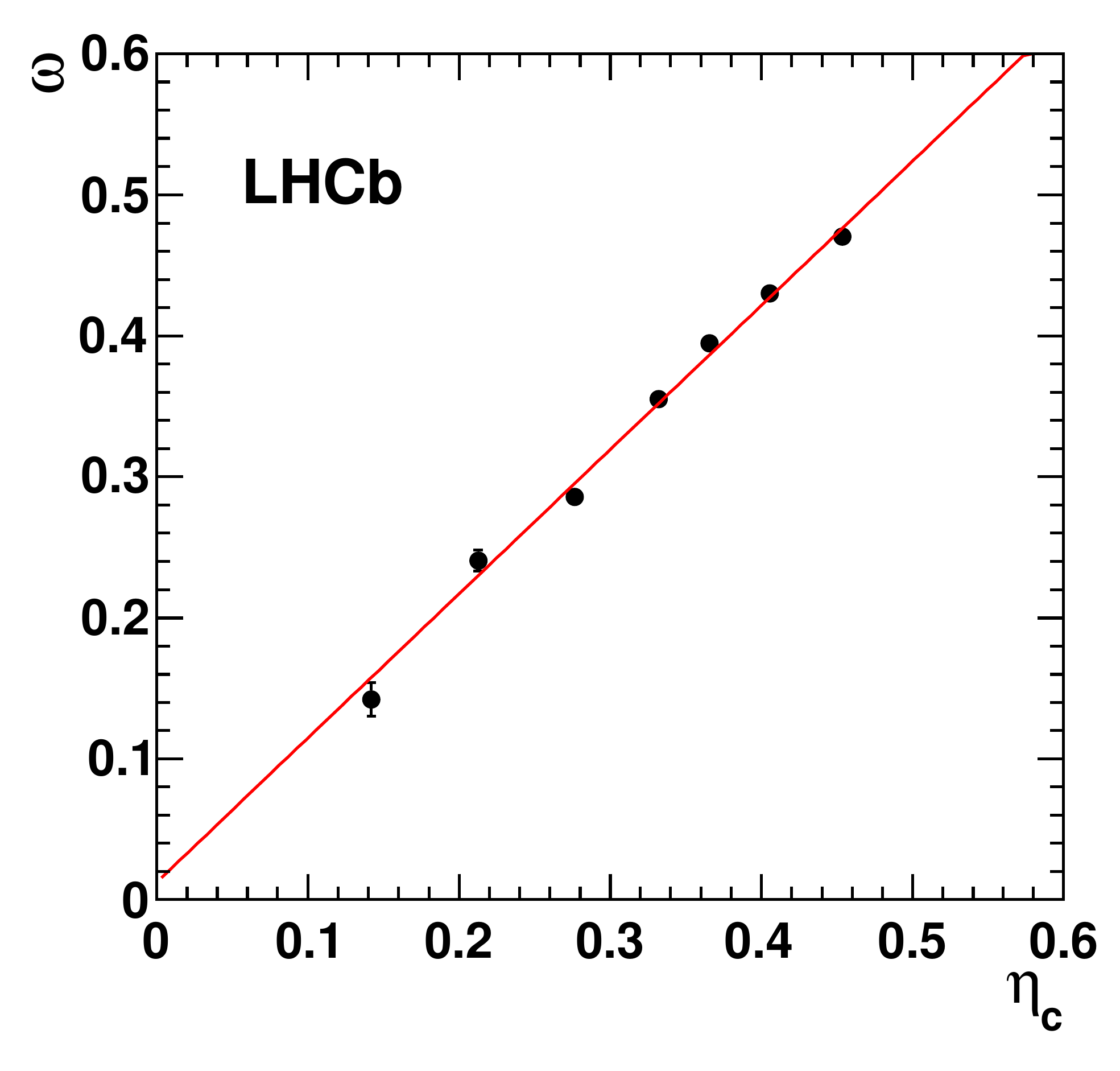}
\end{center}
\caption{\small{Measured mistag fraction (\mistag) versus calculated mistag  
probability ($\eta_c$) calibrated on \bplus signal events for the OS tagger, in 
background subtracted events. 
Left and right plots correspond to \bplus and \dstarmunu signal events.
Points with errors are data, the red lines represent the result of the 
mistag calibration, corresponding to the parameters of Table~\ref{tab:OScal}.
}}
\label{fig:plotCal}
\end{figure}

To extract the calibration parameters in the \bkstar channel 
an unbinned maximum likelihood fit to mass, time and $\eta_c$ is performed. 
In analogy to the \bplus channel, the fit uses the probability density 
functions of $\eta_c$, extracted from data for signal and background 
separately by using the {\it sPlot}~\cite{sWeights} technique.
The results confirm the calibration performed in the 
\bplus channel, albeit with large uncertainties.
The results for the  \dstarmunu channel are obtained from a fit 
to independent samples corresponding to different ranges of the calculated 
mistag probability as shown in Fig.~\ref{fig:dstarmu_oscillation}.
The trigger and offline selections, as well as signal spectra, differ for
this decay channel with respect to the channels containing 
a \jpsi meson. Therefore the agreement in the resulting parameters is 
a validation of the calibration and its applicability to
$B$ decays with different topologies.
In Fig.~\ref{fig:plotCal} the dependency of the measured OS mistag fraction 
as a function of the mistag probability is shown for the \bplus 
and \dstarmunu signal events.
The superimposed linear fit corresponds to the parametrization of 
Eq.~\ref{eq:calibration} and the parameters of Table~\ref{tab:OScal}.
%=======================================================================
%=======================================================================
%==========================================================================

\begin{sidewaysfigure}[h]
	\begin{minipage}[b]{0.32\linewidth}
	\centering
	\begin{overpic}[angle=0, width=0.99\textwidth]{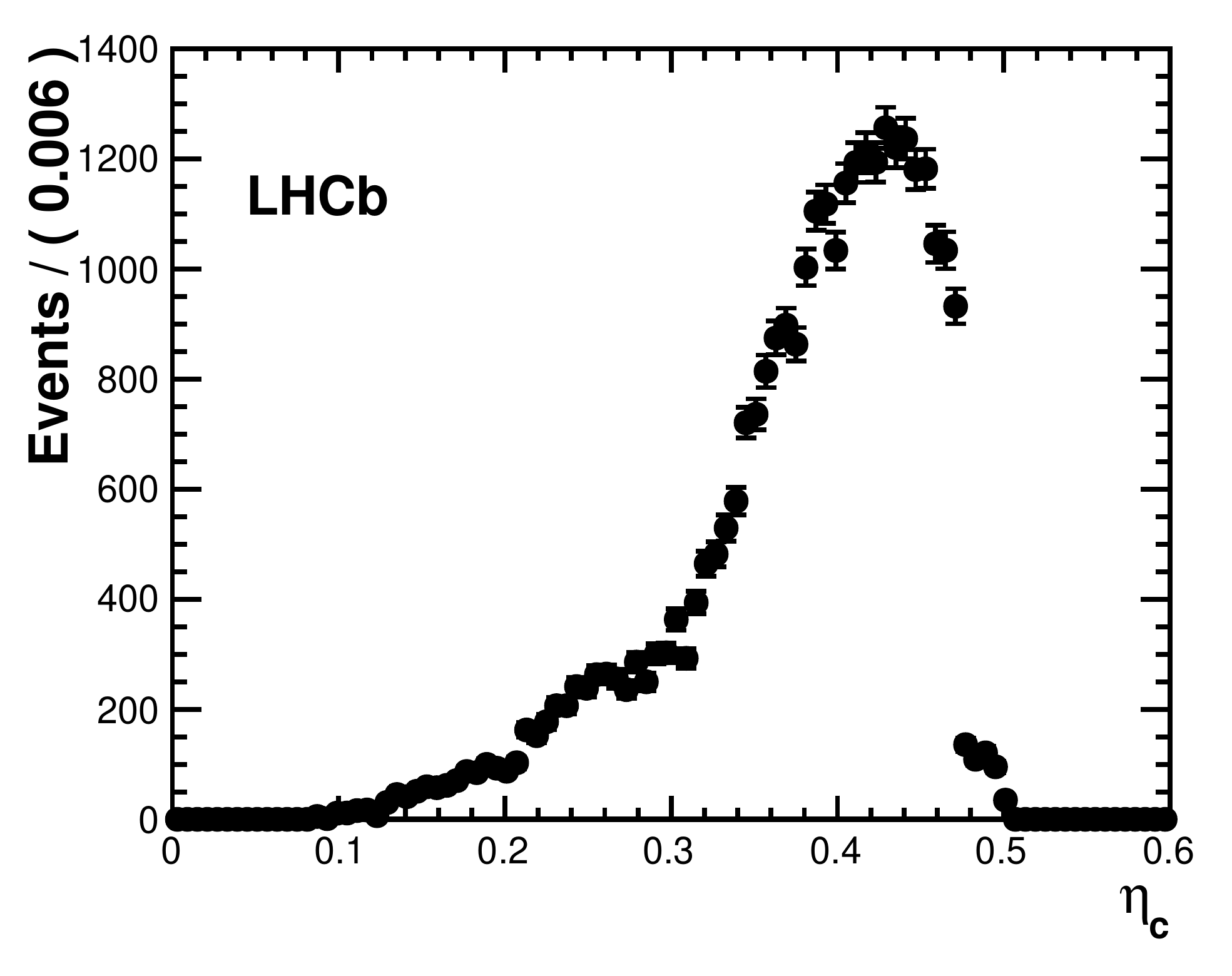}    
	\put (85,68) {\small{(a)}}		
	\end{overpic}
	\end{minipage}
	\begin{minipage}[b]{0.32\linewidth}
	\centering
	\begin{overpic}[angle=0, width=0.99\textwidth]{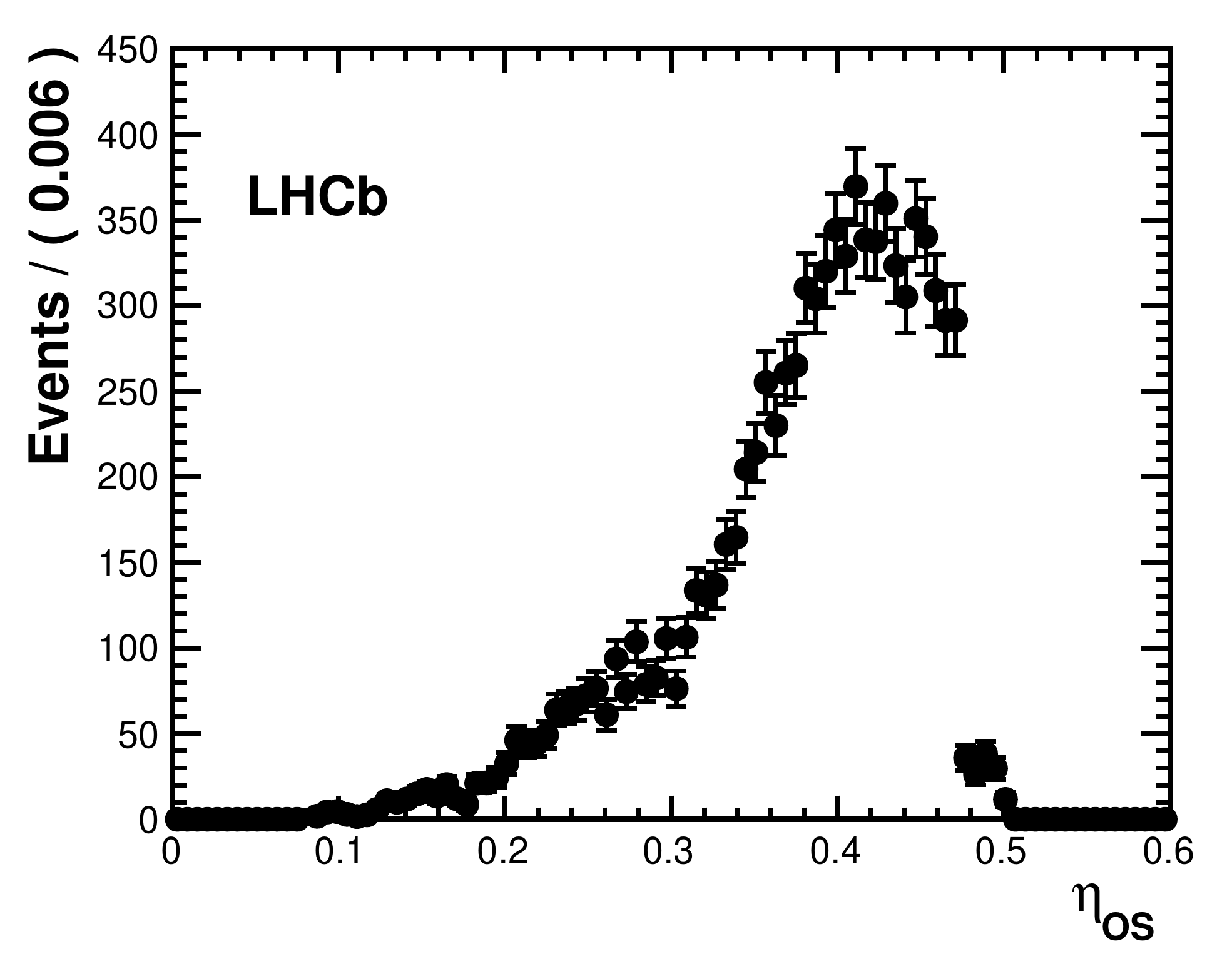}    
	\put (85,68) {\small{(b)}}		
	\end{overpic}
	\end{minipage}
	\begin{minipage}[b]{0.32\linewidth}
	\centering
	\begin{overpic}[angle=0, width=0.99\textwidth]{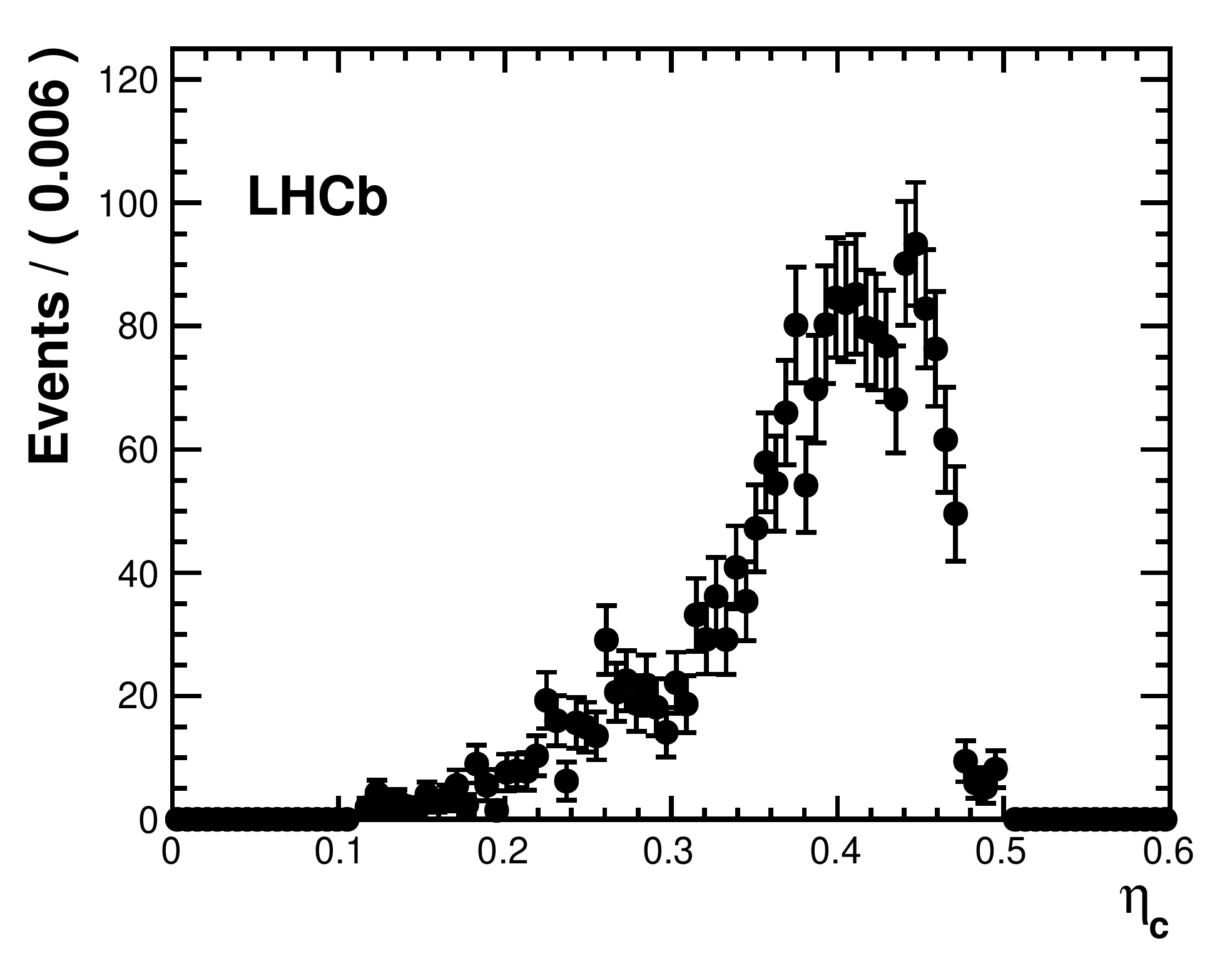}
	\put (85,68) {\small{(c)}}		
	\end{overpic}
	\end{minipage} \\
	\begin{minipage}[b]{0.32\linewidth}
	\centering
	\begin{overpic}[angle=0, width=0.99\textwidth]{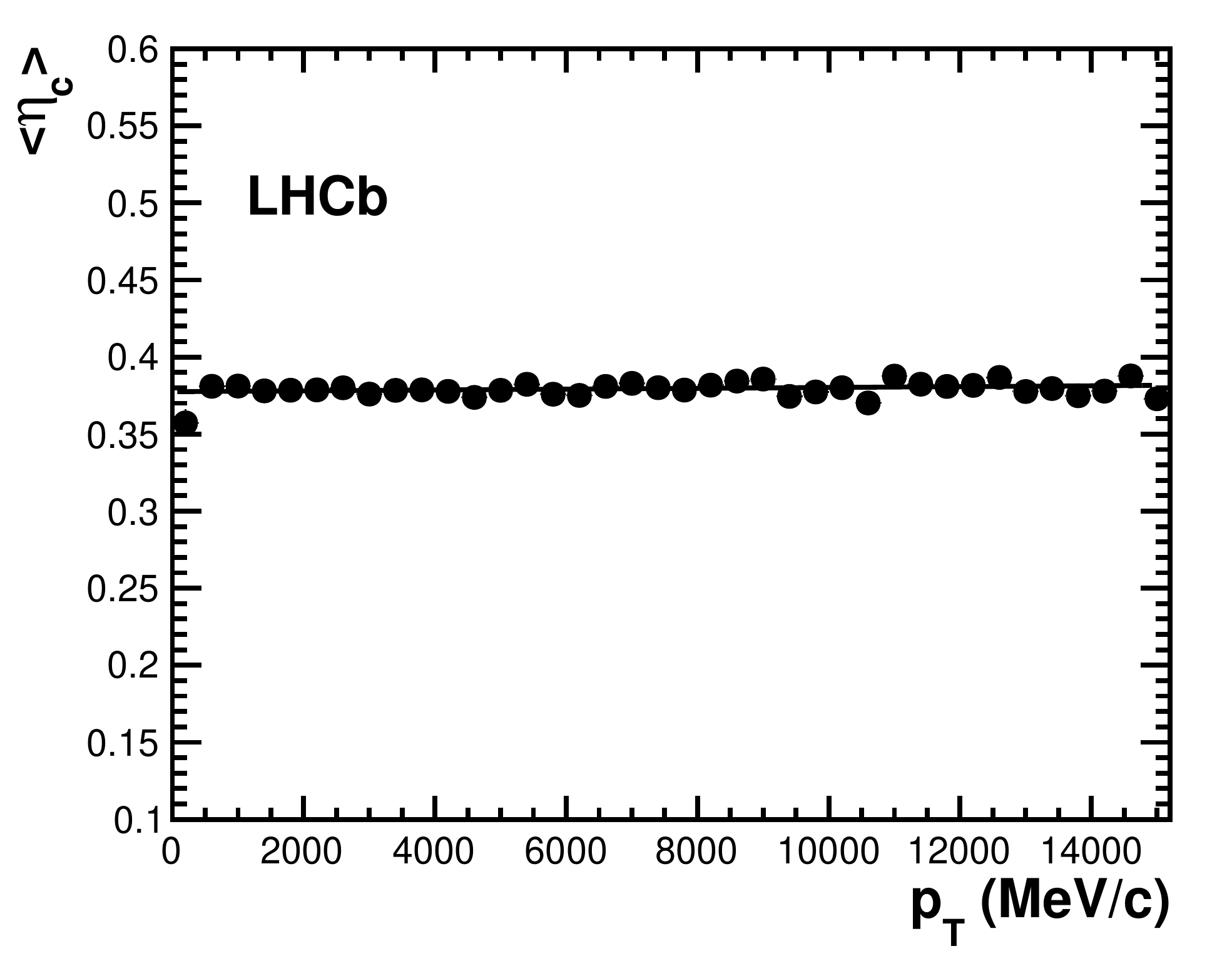}
	\put (85,68) {\small{(d)}}		
	\end{overpic}
	\end{minipage}
	\begin{minipage}[b]{0.32\linewidth}
	\centering
	\begin{overpic}[angle=0, width=0.99\textwidth]{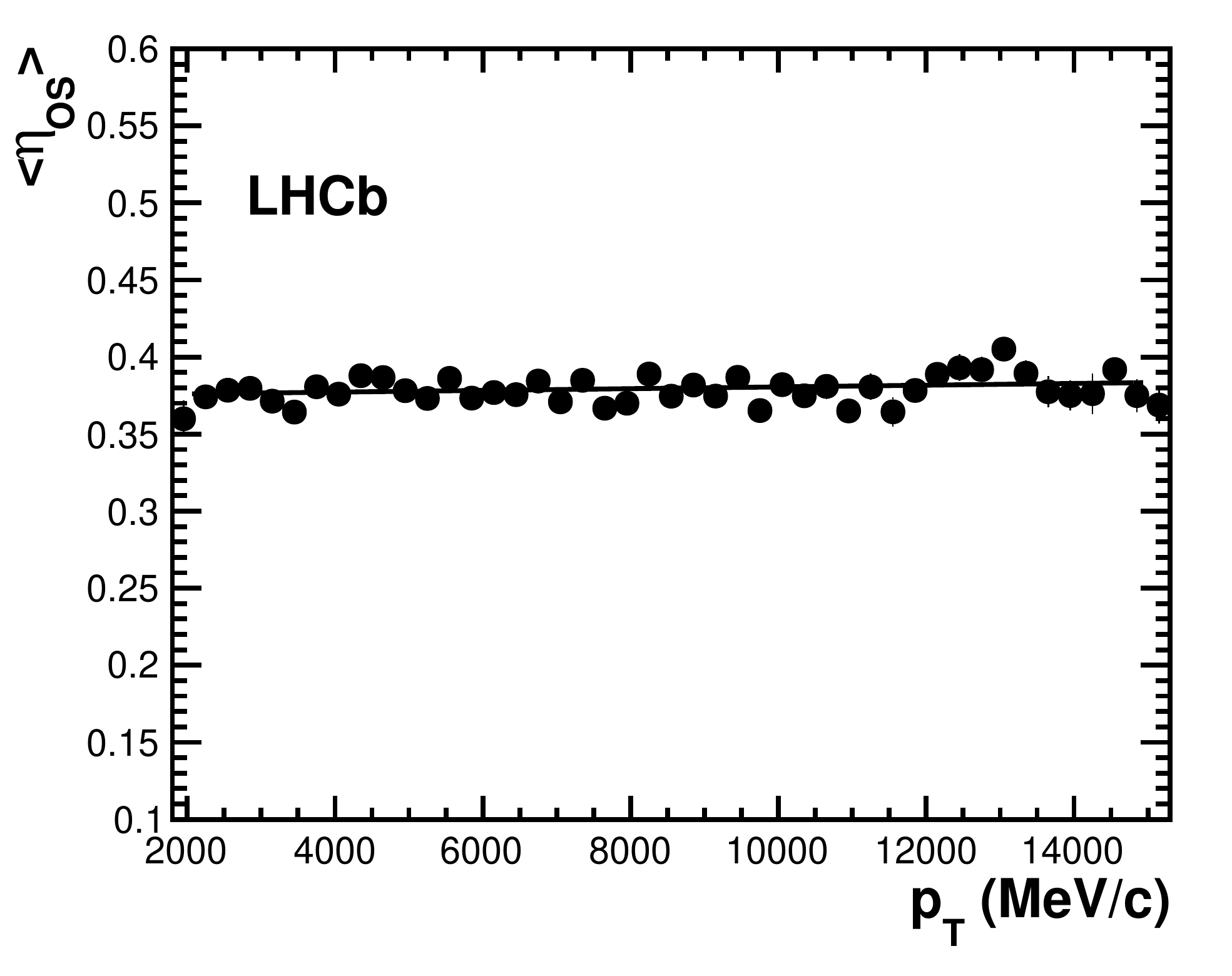}
	\put (85,68) {\small{(e)}}		
	\end{overpic}
	\end{minipage}
	\begin{minipage}[b]{0.32\linewidth}
	\centering
	\begin{overpic}[angle=0, width=0.99\textwidth]{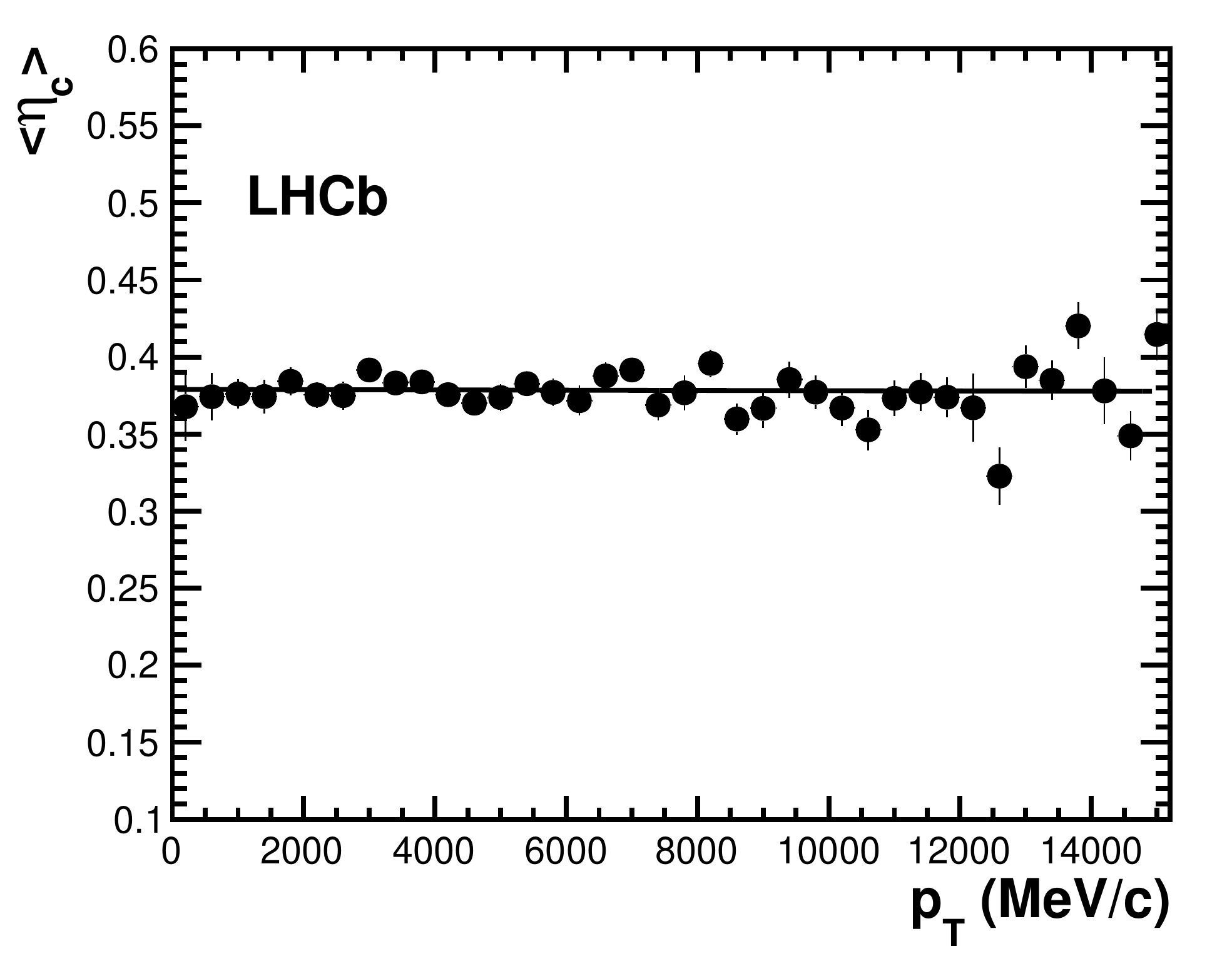}    
	\put (85,68) {\small{(f)}}		
	\end{overpic}
	\end{minipage}
        \caption{\small{Top: calibrated mistag probability distribution for
(a) \bplus, (b) \bkstar and (c) \bs  events.
Bottom: distributions of the mean calibrated OS mistag probability as a 
function of signal \pt for the (d) \Bu, (e) \Bz and (f) \Bs channels. 
The plots show signal events extracted with the {\it sPlot} technique
and with the requirement $t>0.3$\ps.
The three \pt distributions are fitted with straight lines and the slopes are 
compatible with zero.}}
\label{fig:comparison_mistag}
\end{sidewaysfigure}

The output of the calibrated flavour tagging algorithms will be used in a large variety of time-dependent asymmetry measurements, involving different \B decay channels. 
Figure~\ref{fig:comparison_mistag} shows the calculated mistag distributions 
in the \bplus, \bkstar and \bs channels.
These events are tagged, triggered by 
the ``lifetime unbiased'' lines and have an imposed cut of $t>0.3\ps$.
The event selection for the decay \bs is described elsewhere~\cite{phis_phi}.
The distributions of the calculated OS mistag fractions are similar among the channels and the average does not depend on the \pt of the \B.
It has been also checked that the mistag probability does not depend on the 
signal \B pseudorapidity.

%\clearpage

\section{Event-by-event results}
In order to fully exploit the tagging information in the \CP\ asymmetry 
measurements,
the event-by-event mistag probability is used to weight the events accordingly. 
The effective efficiency is calculated by summing the 
mistag probabilities on all signal events
$ \sum_i {(1-2 \omega( \eta^i_c)^2 ) }  /N$.
We underline that the use of the per-event
mistag probability allows the effective efficiency to be calculated on any set 
of selected events, also for 
non flavour-specific channels.
Table~\ref{tab:tagsummary} reports the event-by-event tagging power 
obtained using the calibration parameters determined with the \bplus events
as reported in Table~\ref{tab:OScal}.
The uncertainties are obtained by propagating the statistical and systematic 
uncertainties of the calibration parameters.
In addition to the values for the three control channels the result obtained
for \bs\ events is shown.
For all channels the signal is extracted using the  {\it sPlot} technique.
The results for the tagging power are compatible among the channels containing
 a \jpsi\ meson. The higher value for \dstarmunu is related to the 
higher tagging efficiency.

\begin{table}
\begin{center}
\caption{Tagging efficiency, mistag probability and tagging power calculated 
from event-by-event probabilities for \bplus, 
\bkstar, \dstarmunu and \bs signal events.
The quoted uncertainties are obtained propagating the statistical (first) and 
systematic (second) uncertainties on the calibration parameters
determined from the \bplus events.
}
\begin{tabular}{c|c|c|c}
 \hline
 Channel &   $\etag$ [\%]  &  $\mistag\,$  [\%]  &  $\etag {\cal D}^2 $  [\%] \\ \hline
{\small \bplus } & {\small$27.3 \pm 0.1$  } & {\small $36.1 \pm 0.3 \pm 0.8$ } & {\small $2.10 \pm 0.08 \pm 0.24$ } \\  
{\small \bkstar } & {\small$27.3 \pm 0.3$  } & {\small $36.2 \pm 0.3 \pm 0.8$ } & {\small $2.09 \pm 0.09 \pm 0.24$ } \\ 
{\small \dstarmunu } & {\small$30.1 \pm 0.1$  } & {\small $35.5 \pm 0.3 \pm 0.8$ } & {\small $2.53 \pm 0.10 \pm 0.27$ } \\ 
{\small \bs} & {\small$24.9 \pm 0.5$  } & {\small $36.1 \pm 0.3 \pm 0.8$ } & {\small $1.91 \pm 0.08 \pm 0.22 $ } \\  
\hline
\end{tabular}
\end{center}
\label{tab:tagsummary}
\end{table}

\section{Summary}
Flavour tagging algorithms were developed for the 
measurement of time-dependent asymmetries at the LHCb experiment.
The opposite-side algorithms rely on the pair production
of $b$ and $\bar b$ quarks and infer the flavour of the signal $B$ meson 
from the identification of the flavour of the other 
$b$ hadron. They use the charge of the lepton ($\mu$, $e$) 
from semileptonic \B decays, the charge of the kaon from the $b\to c\to s$ 
decay chain or the charge of the inclusive secondary vertex 
reconstructed from $b$-hadron decay products.
The decision of each tagger and the probability of the decision to 
be incorrect are combined into a single opposite side decision and 
mistag probability.
The use of the event-by-event mistag probability fully exploits 
the tagging information and estimates the tagging power also in 
non flavour-specific decay channels.

The performance of the flavour tagging algorithms were measured on data using 
three flavour-specific decay modes \bplus, \bkstar and \dstarmunu.
The \bplus channel was used to optimize the tagging power and to calibrate 
the mistag probability.
The calibration parameters measured in the three channels are compatible 
within two standard deviations.

By using the calibration parameters determined from \bplus events the 
OS tagging power was determined to be 
$\etag (1-2\omega)^2$ = (2.10$\pm$0.08$\pm$0.24)\% in the \bplus channel, 
(2.09$\pm$0.09$\pm$0.24)\% in the \bkstar channel and 
(2.53$\pm$0.10$\pm$0.27)\% in the \dstarmunu channel, 
where the first uncertainty is statistical and the second is systematic.
The evaluation of the systematic uncertainty is currently limited by the 
size of the available data sample.

\section*{Acknowledgements}

\noindent We express our gratitude to our colleagues in the CERN accelerator
departments for the excellent performance of the LHC. We thank the
technical and administrative staff at CERN and at the LHCb institutes,
and acknowledge support from the National Agencies: CAPES, CNPq,
FAPERJ and FINEP (Brazil); CERN; NSFC (China); CNRS/IN2P3 (France);
BMBF, DFG, HGF and MPG (Germany); SFI (Ireland); INFN (Italy); FOM and
NWO (The Netherlands); SCSR (Poland); ANCS (Romania); MinES of Russia and
Rosatom (Russia); MICINN, XuntaGal and GENCAT (Spain); SNSF and SER
(Switzerland); NAS Ukraine (Ukraine); STFC (United Kingdom); NSF
(USA). We also acknowledge the support received from the ERC under FP7
and the Region Auvergne.

\clearpage
%\input{conclusion}

%\addcontentsline{toc}{section}{References}
%\input{references}
\bibliographystyle{LHCb}
\bibliography{referencesBibTex}
%\bibliography{main}

%\input{lhcb-symbols-list}

\end{document}